\documentclass[letterpaper,twocolumn,10pt]{article}
\usepackage{usenix}

\usepackage{graphicx}
\usepackage{xcolor}
\usepackage{xspace}
\usepackage{enumitem}
\usepackage{tikz}
\usepackage{hyperref}
\usepackage{url}

\usepackage{breakurl}

\usepackage{booktabs}
\usepackage{multirow}
\usepackage{adjustbox}
\usepackage{colortbl}
\usepackage{cleveref}

\usepackage{siunitx}
\usepackage{listings}

\usepackage{tikz-cd}
\usepackage{amssymb}
\usepackage{adjustbox}

\usepackage{soul}

\usepackage{longtable}

\newcommand{\codename}{{\sc Heckler}\xspace}

\renewcommand{\paragraph}[1]{{\bf{\noindent #1}} }
\newcommand{\todo}[1]{\textcolor{red}{TODO: #1}}

\newcommand{\loc}{LoC\xspace}
\newcommand{\eax}{\texttt{eax}\xspace}
\newcommand{\ebx}{\texttt{ebx}\xspace}
\newcommand{\ecx}{\texttt{ecx}\xspace}
\newcommand{\edx}{\texttt{edx}\xspace}
\newcommand{\ebp}{\texttt{ebp}\xspace}

\newcommand{\rax}{\texttt{rax}\xspace}
\newcommand{\rbx}{\texttt{rbx}\xspace}
\newcommand{\rcx}{\texttt{rcx}\xspace}
\newcommand{\rdx}{\texttt{rdx}\xspace}

\newcommand{\rip}{\texttt{rip}\xspace}

\newcommand{\inteighty}{int 0x80\xspace}
\newcommand{\intzero}{int 0x0\xspace}

\newcommand{\etal}{et al.\xspace}

\newcommand{\openssh}{OpenSSH\xspace}

\newcommand{\nx}{\texttt{nx}\xspace}

\newcommand{\authpasswd}{\texttt{auth\_password}\xspace}
\newcommand{\pamsmauth}{\texttt{pam\_sm\_authenticate}\xspace}
\newcommand{\blankpwd}{\texttt{\_unix\_blankpasswd}\xspace}
\newcommand{\mmanspwd}{\texttt{mm\_answer\_authpassword}\xspace}
\newcommand{\auth}{\texttt{auth}\xspace}
\newcommand{\checkf}{\texttt{check}\xspace}

\newcommand{\tapp}{$PT_{\tt{app}}$\xspace}

\newcommand{\ptvm}{$PT_{\tt{vm}}$\xspace}
\newcommand{\pagea}{$P^{\tt{a}}$\xspace}
\newcommand{\pagec}{$P^{\tt{c}}$\xspace}
\newcommand{\attackseq}{$A_{\tt{seq}}$\xspace}

\newcommand{\sboot}{$\mathbb{S}_{\tt{boot}}$\xspace}
\newcommand{\suser}{$\mathbb{S}_{\tt{user}}$\xspace}
\newcommand{\susern}{$\mathbb{S}_{\tt{user}_{i}}$\xspace}
\newcommand{\svm}{$\mathbb{S}_{\tt{vm}}$\xspace}
\newcommand{\sapp}{$\mathbb{S}_{\tt{app}}$\xspace}
\newcommand{\fapp}{$f_{\tt{app}}$\xspace}
\newcommand{\fsudo}{$f_{\tt{app}}$\xspace}

\newcommand{\sbootcardinality}{$|\mathbb{S}_{\tt{boot}}|$\xspace}

\newcommand{\susercardinality}{$|\mathbb{S}_{\tt{user}}|$\xspace}

\newcommand{\sappcardinality}{$|\mathbb{S}_{\tt{app}}|$\xspace}

\newcommand{\tappcardinality}{$|PT_{\tt{app}}|$\xspace}

\newcommand{\ponepam}{$P_{1}^{\tt{pam}}$\xspace}
\newcommand{\ptwopam}{$P_{2}^{\tt{pam}}$\xspace}

\newcommand{\ponesudo}{$P_{1}^{\tt{sudo}}$\xspace}
\newcommand{\ptwosudo}{$P_{2}^{\tt{sudo}}$\xspace}

\newcommand{\ponesshd}{$P_{1}^{\tt{ssh}}$\xspace}
\newcommand{\ptwosshd}{$P_{2}^{\tt{ssh}}$\xspace}

\newcommand{\ponemlp}{$P_{1}^{\tt{mlp}}$\xspace}
\newcommand{\ptwomlp}{$P_{2}^{\tt{mlp}}$\xspace}
\newcommand{\pthreemlp}{$P_{3}^{\tt{mlp}}$\xspace}

\newcommand{\kernack}{\texttt{kern\_ack}\xspace}

\newcommand{\spacingtable}{\vspace{5pt}}

\newcommand{\spacesave}{}

\newcommand{\lessspacesave}{}

\newcommand{\paraspacesave}{}

\newcommand{\circled}[1]{\textcircled{\raisebox{-0.9pt}{#1}}}

\definecolor{c1}{RGB}{207, 34, 46}
\definecolor{c2}{RGB}{5, 80 , 174}
\definecolor{c3}{RGB}{36, 41, 47}

\definecolor{codegreen}{rgb}{0,0.6,0}
\definecolor{codegray}{rgb}{0.5,0.5,0.5}
\definecolor{codepurple}{rgb}{0.58,0,0.82}
\definecolor{textcolor}{rgb}{0.3,0.3,0.3}
\usepackage{bold-extra}

\definecolor{backcolour}{rgb}{0.95,0.95,0.92}

\definecolor{rulecolor}{rgb}{0.87,0.87,0.87}

\lstdefinestyle{mystyle}{
    keywordstyle=\color{c2}\bfseries,
    numberstyle=\tiny\color{codegray},
    commentstyle=\color{gray},
    basicstyle=\linespread{1.3}\color{c3}\ttfamily\scriptsize\bfseries,
    breakatwhitespace=false,         
    breaklines=true,                 
    captionpos=b,                    
    keepspaces=true,                 
    numbers=left,                    
    numbersep=5pt,                  
    showspaces=false,                
    showstringspaces=false,
    showtabs=false,                  
    tabsize=2,
    frame=single,   
    rulecolor=\color{rulecolor},
    framexrightmargin=-2pt,
    framexbottommargin=0pt,
    framextopmargin=0pt,
}

\lstdefinelanguage{Julia}%
  {morekeywords={abstract,break,case,catch,const,continue,do,else,elseif,%
      end,export,false,for,function,immutable,import,importall,if,in,isa,rethrow,%
      macro,module,otherwise,quote,return,switch,true,try,type,typealias,%
      using,while},%
   sensitive=true,%
   alsoother={\$},%
   morecomment=[l]\#,%
   morecomment=[n]{\#=}{=\#},%
   morestring=[s]{"}{"},%
   morestring=[m]{'}{'},%
}[keywords,comments,strings]%

\lstdefinelanguage
   [heckler]{Assembler}     %
   [x86masm]{Assembler} %
   {morekeywords={CDQE,CQO,CMPSQ,CMPXCHG16B,JRCXZ,LODSQ,MOVSXD, %
                  POPFQ,PUSHFQ,SCASQ,STOSQ,IRETQ,RDTSCP,SWAPGS, %
                  rax,rdx,rcx,rbx,rsi,rdi,rsp,rbp, %
                  r8,r8d,r8w,r8b,r9,r9d,r9w,r9b, %
                  r10,r10d,r10w,r10b,r11,r11d,r11w,r11b, %
                  r12,r12d,r12w,r12b,r13,r13d,r13w,r13b, %
                  r14,r14d,r14w,r14b,r15,r15d,r15w,r15b},
                    sensitive=true,%
   alsoother={\$},%
   morecomment=[l]\%,%
   morecomment=[n]{\#=}{=\#},%
   morestring=[s]{"}{"},%
   morestring=[m]{'}{'},
                 }[keywords,comments,strings]%

\crefname{figure}{Fig.}{Figs.}
\crefname{section}{Sec.}{Secs.}
\crefname{appendix}{Appx.}{Appx.}
\crefname{table}{Tab.}{Tab.}

\usepackage{amsthm}

\theoremstyle{definition}

\newcommand{\Spurious}{{Malicious}\xspace}
\newcommand{\spurious}{{malicious}\xspace}

\newcommand{\sevstep}{SEV-Step\xspace}

\begin{document}

\date{}

\title{\codename: Breaking Confidential VMs with \Spurious Interrupts
}

\author{\rm Benedict Schlüter \qquad Supraja Sridhara \qquad Mark Kuhne \qquad Andrin Bertschi \qquad Shweta Shinde  \\
{ETH Zurich}
} 

\maketitle
\begin{abstract}
\vspace{5pt}
Hardware-based Trusted execution environments (TEEs) offer 
an isolation granularity of virtual machine abstraction. They provide confidential VMs (CVMs) that host security-sensitive code and data. 
AMD SEV-SNP and Intel TDX enable CVMs and are now available on popular cloud platforms.
The untrusted hypervisor in these settings is in control of several resource management and configuration tasks, including interrupts. 
We present \codename, a new attack wherein the hypervisor injects malicious non-timer interrupts to break the confidentiality and integrity of CVMs. 
Our insight is to use the interrupt handlers that have global effects, such that we can manipulate a CVM's register states to change the data and control flow.
With AMD SEV-SNP and Intel TDX, we demonstrate \codename on \openssh and sudo to bypass authentication. On AMD SEV-SNP we 
break execution integrity of C, Java, and Julia applications that perform statistical and text analysis.
We explain the gaps in current defenses and outline guidelines for future defenses.
\footnotetext[0\def\thefoornote{}]{This is the author's version of the USENIX Security 2024 paper.}

\end{abstract}

\pagestyle{plain} %

\section{Introduction}

Hardware-based trusted execution environments (TEEs) flip the conventional trust mode. They designate the cloud service provider and privileged software such as the hypervisor as untrusted entities.
Recent TEEs lean towards a virtual machine abstraction for isolation granularity to provide {\em confidential VMs (CVMs)} that host security-sensitive code and data. 
AMD Secure Encrypted Virtualization-Secure Nested Paging (SEV-SNP) and Intel Trust Domain Extensions (TDX) are the two main extensions offered currently from hardware providers~\cite{sev-snp,TDX}, while Arm Confidential Computing Architecture (CCA) is anticipated to be in production in the future~\cite{cca}. 
CVMs have received wide-scale adoption as cloud confidential computing hosted by major cloud providers such as Google Cloud, Microsoft Azure, Alibaba Cloud, and IBM Cloud~\cite{google-tdx, azure-sev-tdx, google-sev, ibm_confidential_cloud, alibaba-tdx}.

Hardware isolation and memory encryption in TEEs ensure the confidentiality and integrity of CVMs.
However, despite being untrusted, the privileged software components such as the hypervisor remain responsible for resource allocation and virtualization management. 
As a result, it's crucial to reconsider how these untrusted components interact with the CVMs.
We examine one such class of interfaces, namely the interrupt management that is under the hypervisor's control.

In this paper we present \codename, a new software-based attack that breaks the confidentiality and integrity of CVMs on AMD SEV-SNP and Intel TDX.
\codename leverages the untrusted hypervisor's ability to inject controlled interrupts into the victim CVM at points of its choice. 
Since the CVMs run a full-fledged trusted operating system, it has valid handlers for several interrupts.
Thus, unbeknownst to itself, the victim CVM starts executing the interrupt handlers corresponding to the interrupt injected by the hypervisor. 
However, unlike timer interrupts that are widely used for side-channel attacks~\cite{xu2015controlled, 203868, sgx-step, nemesis} because of their effects on cache and micro-architectural states, the CVM has handlers change registers and global state thus impacting the subsequent execution. 
Thus by simply injecting interrupts, the hypervisor is able to change the victim VM's data and control flow. 

\codename is part of a larger family of attacks where an privileged attacker sends malicious notifications to the victim running in a TEE. 
We coin the term {\em Ahoi} to refer to this class of attacks.\footnote{Ahoi is a signal word to call a ship or boat.
It is also an anagram of Iago~\cite{checkoway2013iago} with edit distance of one.} 
Previous studies that exploit timer interrupts and page faults fall within the category of Ahoi attacks---they produce malicious interrupts to allow the attacker to monitor side-effects like cache and timing. 
Unlike these prior instances of Ahoi, \codename generates interrupts that go beyond side-effects; it targets explicit effect handler execution that directly modifies registers i.e., the CVM's global state.

\paragraph{Findings.}
We analyze the hypervisor's interrupt injection behavior on AMD SEV-SNP and Intel TDX.
We find that both of them forward some, if not all, interrupts to the victim CVMs.
Notably, both of them allow the attacker to inject \inteighty on cores executing CVMs. %
As an effect, the CVM executes the corresponding handler 
on behalf of a user-space process (e.g., statistical analysis, user authentication, daemons) that is currently executing on the core. 
Worse yet, as per the semantics of \inteighty, the handler treats the current register state set up by the process as syscall number (\rax) and input args (\rbx, \rcx, \rdx) for the system call.
The guest kernel in the CVM, completely unaware that the hypervisor and not the process invoked this handler, executes the system call and returns the result of the system call back to the process by updating its \rax. 
\codename abuses this behavior to operate as a gadget that changes the victim programs' \rax. 
Further, AMD SEV-SNP allows the attacker to inject other interrupts such as \intzero and many more. Some of these interrupts are presented as signals to the user program. 
We find that the application-specific handler for these signals can have global side effects. For example, scientific calculations 
have handlers to convert the operands of faulting instructions  
(e.g., the denominator in a divz is set to a NaN) to capture specific notions (e.g., $\infty$, -$\infty$). \codename changes this behavior into a gadget to convert particular program variables (e.g., to NaN) and continue execution. 
Lastly, we can chain gadgets by injecting multiple interrupts at selective locations of victim's execution to change more than one data and control flow.

\paragraph{Orchestrating \codename.}
End-to-end exploits built with \codename require injecting interrupts at targeted execution points in the victim programs to induce effects brought on by our gadgets. 
Specifically, we need to know the exact core on which the user program executes inside the CVM, the guest physical address of the point of gadget injection, and the moment when the program reaches the point of interest in its execution. 
For AMD SEV-SNP, we use several heuristics particular to our target programs based on the information we can glean about its execution (e.g., page faults). 
We maximize this by leveraging auxiliary information leaked by observable behavior despite encryption of CVM state (e.g., order of page accesses, execution in shared libraries)~\cite{SEVered, sev-step}.

\paragraph{Implications.}
We use the \codename gadgets to alter the data and control flow of five case-studies to break confidentiality and integrity of CVMs.
First, on AMD SEV-SNP and Intel TDX, we bypass the authentication in \openssh and sudo, thus allowing the hypervisor to gain complete root access to the CVM. 
Next, we break execution integrity of AMD SEV-SNP by altering the 
results of statistical and text analysis in C, Java, and Julia.
Lastly, we discuss the effectiveness of existing defenses offered by AMD SEV-SNP and show that they are insufficient. We develop kernel-patches for Intel TDX to stopgap the effects of our \inteighty gadget.

\paragraph{Contributions.}
We make the following novel contributions:

\begin{itemize}
\itemsep0em
\lessspacesave
\vspace{-5pt}
\item {Novel Attack.}
We introduce \codename, a new attack wherein a hypervisor injects malicious interrupts to trigger handlers that change the data and control flow of victim CVMs.
\item {Gadgets \& Chaining.} %
We identify several crucial gadgets in prevalent services and workloads typically hosted in cloud-based CVMs. We invoke and chain these gadgets using custom orchestration techniques. 
\item {Proof-of-concept Exploits.} 
We show that our AMD SEV-SNP and Intel TDX exploits can bypass \openssh and sudo;
our AMD SEV-SNP exploits can break statistical and text analysis for AMD SEV-SNP. This demonstrates that \codename breaks the integrity and confidentiality guarantees offered by these state-of-the-art TEEs.
\end{itemize}

\paragraph{Disclosure.} 
We informed Intel and AMD about \inteighty on 27 and 28 September 2023 respectively. 
We updated AMD on 14 October 2023 about our findings for other interrupts and our analysis of their defenses. 
\codename is tracked under two CVEs: 
CVE-2024-25744 for \inteighty was mitigated with a kernel patch for SEV-SNP and TDX~\cite{int80-path-1}.
CVE-2024-25743 for other interrupts remains unmitigated for AMD on 6 March 2024 at the time of the writing.

\noindent 
\codename tooling and PoC exploits are open-source at: \\ \url{https://ahoi-attacks.github.io/heckler}

\spacesave
\section{Overview}
\label{sec:ovw}
\spacesave

Hardware-based trusted execution environments provide an abstraction to execute code and data, such that its confidentiality and integrity is preserved even in the presence of privileged software. 
AMD Secure Encrypted Virtualization-Secure Nested Paging (AMD SEV-SNP), AMD Secure Encrypted Virtualization-Encrypted State (SEV-ES), and Intel Trust Domain Extensions (Intel TDX) provide a VM-level abstraction called confidential VMs (CVMs). 
For these TEE abstractions, the untrusted privileged hypervisor provisions the execution resources (e.g., CPU and memory) for VMs.  
The hardware ensures execution and memory isolation such that the untrusted software cannot compromise the CVM.

Notably, the untrusted hypervisor provides virtualization abstractions such as interrupt routing to CVMs.
Thus, the attacker can abuse this interface to inject non-genuine (e.g., wrong interrupt number) and unexpected interrupts (e.g., at the wrong instruction), i.e., {\em \spurious interrupts} into the target. 
Physical timers, the most widely-studied interrupt, have 
been shown to break the confidentiality of TEEs 
by amplifying side-channel attacks~\cite{sgx-step}.
However, other interrupts have received little to no attention, 
because they are assumed to never explicitly affect the victim's execution beyond side-effects that can be gleaned via side-channels. 

\spacesave
\subsection{Interrupt Delivery to CVMs}
\label{sec:interrupt-delivery}

\begin{figure}[t]
    \centering
     \includegraphics[scale=0.65]{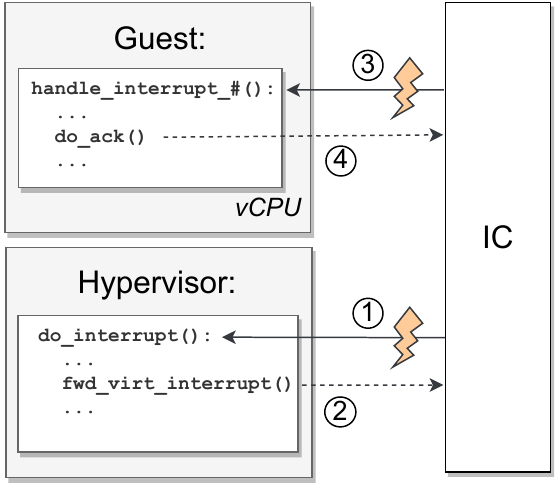} 
         \caption{Virtualized interrupt for CVMs.
         Solid arrows (\circled{1}, \circled{3}): asserted interrupt lines; dotted arrows (\circled{2}, \circled{4}): memory-mapped write. The interrupt controller (IC) delivers a physical interrupt to the hypervisor \circled{1}. The hypervisor writes to a memory-mapped region of memory \circled{2} that emulates a virtual Interrupt Controller (vIC) for the vCPU to forward the virtual interrupt \circled{3}. The OS writes to a memory-mapped register in the vIC to acknowledge the interrupt \circled{4}. 
         } 
    \label{fig:virt-int}
 \end{figure}

The guest OS executing inside the CVMs relies on interrupts for its operation (e.g., the Linux kernel requires timer interrupts for scheduling).
Therefore, similar to traditional virtualization in non-confidential execution, the hypervisor has to virtualize the interrupt management and delivery to the CVMs. 
To do so, the hypervisor hooks on all physical interrupts in the interrupt controller. 
~\cref{fig:virt-int} shows this mechanism at a high-level. 
For every interrupt, the hypervisor determines which VM the interrupt should be routed to, based on the CPU-to-vCPU mapping it maintains.  
Then, the hypervisor forwards the virtual interrupt to the vCPU.
The guest OS of the CVM services the virtual interrupt.
Finally, the guest OS acknowledges the interrupt in the interrupt service routine (ISR). 
The SEV and TDX hardware implementations and hardened guest Linux images (called enlightened guest OS) attempt to limit the interfaces that a CVM exposes to the untrusted hypervisor. However, our analysis shows that the hypervisor is still able to inject certain or all types of interrupts (see Sec.~\ref{sec:handler-results} for results).

\subsection{\codename Attack}
\label{sec:attack-ovw}
The hypervisor can arbitrarily inject interrupts to the CVMs. 
Such interrupts cause the guest OS to execute its interrupt service routines (ISRs)
which can have side-effects that an attacker can exploit. 
For example, Linux uses interrupt number 0x80 for legacy 32-bit system calls on x86. 
Asserting interrupt 0x80 triggers the corresponding ISR. 
The ISR reads register \eax and executes the system call.
Further, it stores the result of the system call in the \eax register. 
Note that this system call interface only updates the \eax register. 
All other registers are restored by the kernel before returning to the user-space. 
Therefore, a malicious hypervisor can inject interrupt 0x80 and change the value stored in \eax.

\begin{figure}[t]
    \centering
     \includegraphics[scale=0.6]{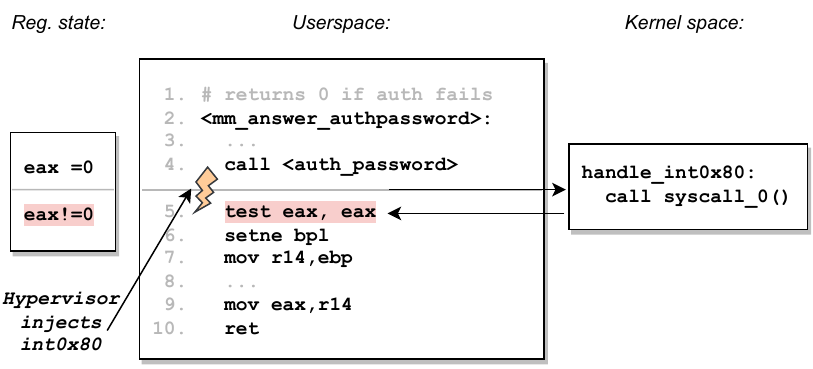} 
         \caption{Inject \inteighty for \openssh authentication. \mmanspwd is invoked during ssh authentication. It returns  $0$ when authentication fails. A malicious \inteighty triggers a call to the syscall $0$ handler which sets \eax to a non-zero value when \authpasswd returns, resulting in a successful authentication.}
    \label{fig:attack-sshd}
 \end{figure}

\paragraph{Attacking \openssh.}
We consider the \openssh application executing in user-space that runs a server process sshd.
A CVM may host this process to allow trusted users to login and manage the confidential services.
The SSH authentication routine in sshd invokes the \mmanspwd function to check the user's credentials.
If authentication fails, the function returns 0. 
The disassembly of this function shows that the return value of \authpasswd is stored in the \eax register (see Lines  5-10 in \cref{fig:attack-sshd}).
Further, the caller of \mmanspwd checks if the return value is non-zero, and if so, allows the user to login. 
Consider the case where the attacker is trying to log into the CVM. Since it does not have the correct user credentials, the return value of \authpasswd and consequently \mmanspwd will always be 0. 
However, if the attacker can change \eax from zero to a non-zero value, then the caller of \mmanspwd will let the attacker login, despite using wrong credentials. 
From a malicious hypervisor's perspective, if it injects an \inteighty right after the return of \authpasswd, it can indeed change the value of \eax before it is used by \mmanspwd. 
Then, \mmanspwd returns a non-zero value to the caller. 
The only thing that remains is to trigger \inteighty such that it returns some other non-zero value in \eax.
If we take a closer look at the point of interrupt injection, \eax is set to 0 by the function \authpasswd. If a malicious hypervisor injects an \inteighty at this point, it triggers the execution of the handler on behalf of the sshd process. This results in executing system call number 0. 
In the Linux kernel, this corresponds to the restart system call which should always be invoked from within the kernel. 
Since we invoke it from the user-space, the kernel returns an \texttt{EINTR} error ($-4$, i.e., a non-zero value) in \eax. 
In summary, the hypervisor uses the interrupt injection primitive to gain access to the CVM.

\spacesave
\section{\Spurious Interrupts}
\label{sec:spurious-interrupts}
\spacesave
\codename leverages the effects of interrupt handlers on user-level applications, such that the attacker can alter their benign behavior to do its bidding. 
Apart from the \inteighty handler we used in our motivating example, we systematically analyze other interrupts and their potential use in \codename.

\paragraph{Threat Model.}
We operate in the standard threat model of confidential VMs provided by Intel TDX and AMD SEV-SNP. 
The untrusted hypervisor loads the CVM image in memory and controls the initial configurations. Remote attestation measures the CVM's initial memory before initiating the boot up. 
The software executing inside the CVM (guest OS, user applications, trusted modules for TEEs) is included in the TCB.
As for configurations, 
the specifications for TDX and SEV-SNP outline certain initial state that the hypervisor has to setup (e.g., number of vCPUs, supported hardware  features, memory size). The hardware checks this and only enters the CVM if the setup is correct. 
The hardware zeroes out certain values (e.g., certain general-purpose registers) before exiting the CVM. 
During execution, SEV-SNP and TDX encrypt and integrity protect the VM pages. Further, they protect register state and check some control and communications pages (e.g., Virtual Machine Control Block) that are shared with the hypervisor. 
The hypervisor is still expected to manage the CVMs by allocating physical pages and scheduling vCPUs. 
This includes injecting interrupts through different interfaces such that the CVM can continue to perform its tasks (e.g., virtio updates) and to notify the CVM about critical interrupts (e.g., virtual timers).
We note that the specific protections of state shared between the hypervisor and the CVM vary for AMD SEV-ES, AMD SEV-SNP, and Intel TDX.

\paragraph{Scope.} It has been shown that attacking AMD SEV-SNP is more challenging than attacking AMD SEV-ES~\cite{sev-snp}. This is mainly because SEV-ES does not provide integrity protection~\cite{SEVurity}.
We leave attacks on AMD SEV-ES out of scope for this paper and instead focus on AMD SEV-SNP, with the expectation that if the attacks work on SEV-SNP, they will work on SEV-ES as well.

\spacesave
\subsection{Trace-based Reasoning}
\label{sec:trace-reasoning}

Our goal is to identify interrupt handlers that, when executed at arbitrary points during a victim program execution, induce changes that impact the application. 
To capture this systematically, we introduce the notion of traces as defined below. 

\paragraph{Trace.}
Consider a given program P and an input I that produces output O. Then program trace $T_P~(I, O)$ is a sequence of states $S_1, \dots, S_n$, where $S_i$ is the program state that captures registers and virtual memory at time $t_i$. 
\lessspacesave
We capture explicit inputs as well as environment variables in $I$, and our state captures the register states and virtual memory of the process.
Note that for a given $P$, $I$, $O$, its trace $T_P(I, O)$ is always deterministic. 

\paragraph{Explicit Effect Handlers.}
If a program $P$ incurs a fault, interrupt, exception, or signal during its benign execution, then the system executes a handler either in the guest kernel or user space via an application-registered handler.
The trace $T$ captures it gracefully. 
For kernel handlers, they do not affect the program and hence are not accounted for in the trace. 
If the program executes a handler to terminate the program, that is captured by the state with the last state being program exit. 
More importantly, handlers that update the program state and continue execution are also captured by the notion of states. 
For example, consider a program with a custom floating point error handler that rounds off the value to the nearest integer, say 1. 
When the program executing on input $I$ is in state $S_i$, it receives a SIGFPE for an operation on variable $a$ that overflows. The program executes the handler that converts the problematic variable from $a$ to $a'$, thus changing the state from $S_a$ to $S_{a'}$.
We refer to such handlers, that effect a state change, as {\em explicit effect handlers}.
But, if the program receives a timer interrupt then the program states stay unaffected. 

\begin{figure}
  \vspace{-5pt}
   \centering
    \includegraphics[scale=1]{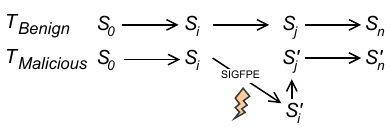} 
    \vspace{-13pt}
\caption{$T_{Benign}$ and $T_{Malicious}$ represent traces for benign and malicious execution of $P$ under input $I$. This leads to traces $S_0, S_i, S_j, \dots, S_n$ and $S_0, S_i, S'_i, S'_j, \dots, S'_n$ to produce outputs $O$ and $O'$ respectively. The attacker injects \intzero when $P$ is in state $S_i$. This induces a state $S'_i: S_i[mem|mem[a]\mapsto 1]$, where the memory that holds variable $a$ (i.e., $mem[a]$) is set to $1$.}
\label{fig:sts-eg}
\end{figure}

\paragraph{Inducing Malicious State Transitions.}
The attacker has the capability to inject arbitrary interrupts into the CVM to invoke the corresponding handlers. For example, consider a benign execution of program $P$. At time $t_i$, it is in state $S_i$ and changes to $S_j$ at $t_{i+1}$. 
However, under a malicious execution, at time $t_i$, the attacker sends an \intzero to the VM's vCPU that is executing $P$ who receives a SIGFPE. 
P's handler will execute at $t_{i+1}$, thus inducing a malicious state transition from $S_i$ to $S'_j$. If we consider our above described handler that sets variable $a$ to 1 on SIGFPE, the attacker has successfully managed to achieve a state transition from $S_i$ to  $S_{j'}$ where $mem[a] \mapsto 1$.
Worse yet, since the handler resumes execution of the program, the attacker can time the interrupt such that the subsequent program logic uses the modified state variables, $a$ in our example, thus leading to a different data or control flow and trace (see~\cref{fig:sts-eg}).

\spacesave
\subsection{Detected Explicit Effect Handlers}
\label{sec:handler-results}
We first analyze the hypervisor's ability to inject interrupts into the CVM, both on Intel TDX and AMD SEV-SNP. 
For this, we conduct a simple test on AMD SEV-SNP and Intel TDX machines (see Sec.~\ref{sec:e2e-attacks} for CPU and software details). We enumerate the interrupts from 0-255, the valid range of interrupts that a VM can receive. 
We inject them in our victim application executing inside the CVM via the hypervisor-provided interface. 
Then, we use $2$ main observations regarding the x86 architecture to detect explicit event handlers for interrupts: (a) it has an explicit instruction that uses the interrupt number 128 (i.e., \texttt{int 0x80}) to perform legacy system calls, and (b) the Linux kernel maps interrupts to signals that are delivered to user-space applications. 
First, we test if \inteighty is delivered to the CVM on both AMD SEV-SNP and Intel TDX machines when injected from the hypervisor. 
We see that the Linux kernel's \inteighty handler always returns the result of the legacy system call in the \texttt{eax} register. Further, the different system call handlers conditionally read \texttt{ebx}, \texttt{ecx}, \texttt{edx}, \texttt{esi}, and \texttt{edi} registers. 

Next, to detect if interrupts from the hypervisor are delivered as signals to the user application, 
we write a C application that registers handlers for all signals and waits in a busy loop.
With this setup, we inject all interrupts to the CVM.
For a given interrupt, if the CVM has a valid handler registered we can observe its impact, if any, on the application. 
We see that, for most interrupts, the Linux kernel uses a default handler that acknowledges the interrupt in the kernel and has no explicit effect on the application. 
Next, we summarize our specific findings for interrupts that impacted the applications. 

\paragraph{SEV.}
Our experiments show that all interrupts were delivered to the CVM and handled by the guest Linux kernel. 
We observe that \inteighty is delivered to the CVM and always noticeably impacts the user application.
Further, the guest Linux kernel delivers $11$ interrupts as a signal to the user-space application. 
Therefore, these $12$ interrupts have explicit effects on the application.

\paragraph{TDX.}
All interrupts below $31$ were dropped by the hardware and never even delivered to the guest VM. The only interrupt that was selectively allowed in this range was an NMI.  
For interrupts above $31$ that reached the guest VM, only \inteighty noticeably impacted the application.

\spacesave
\section{\codename Gadgets}
\spacesave

Next, we detail particular explicit effect handlers we detected and their exact effects. We refer to handler code as a {\em \codename gadget}, inspired by memory corruption attacks~\cite{rop, dop}.

\begin{figure}
    \centering
     \includegraphics[scale=0.8]{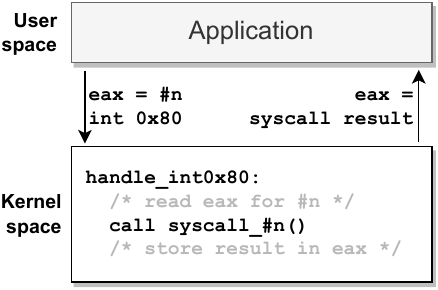} 
         \caption{ For int 0x80, the Linux kernel executes a system call corresponding to the number (\#n) stored in \eax by the application. When returning to the application, the kernel stores the result of the system call in the \eax register. }
    \label{fig:int-80}
 \end{figure}
 
\subsection{Syscalls from Userspace}
Linux uses \inteighty for legacy system calls as shown in Fig.~\ref{fig:int-80}. 
Asserting \inteighty triggers the corresponding ISR in the kernel space of the CVM. 
The ISR reads register \eax and executes the corresponding system call.
Further, it stores the result of the system call in the \eax register.
Therefore, a malicious hypervisor can inject \inteighty and arbitrarily change the value stored in \eax at any time (see~\cref{sec:attack-ovw}). 
Further, based on the value in \eax an attacker can use this interface to execute arbitrary system calls to attack the victim CVM (e.g., change page permissions, copy memory).\footnote{\inteighty instruction can be executed in 64 and 32-bit binaries.}

\paragraph{Example.} Consider an application that stores a secret on the stack (\ebp-4) 
and accesses shared memory in the non-secure region (e.g., for communication with a non-secure VM).
An attacker can use the \inteighty to leak this secret by triggering the write system call.
The Linux kernel executes the write system call in the \inteighty handler (see~\cref{fig:int-80}) when \eax is set to $4$. 
Then, with the right parameters, such a call writes the secret to the hypervisor accessible shared memory.
Specifically, the write system call takes 3 parameters; (fd) a file descriptor to write to in \ebx, (buf) the address to read from in \ecx, and (count) the number of bytes to read in \edx.
Therefore, we need an application that has a gadget as shown in the code snippet below:
\begin{lstlisting}[language={[heckler]Assembler},escapechar=|]
mov eax , 4          %
mov ebx ...          %
mov ecx, [ebp - 4]   %
mov edx, 8           %
... |\label{line:asm:inj}| 
\end{lstlisting}
Now, if the hypervisor injects \inteighty on line~\ref{line:asm:inj}, the kernel in the CVM will execute the \texttt{write} system call and leak the secret in \ecx to the shared memory region in \ebx. 
Note that, this program never executes the \texttt{int 0x80} instruction. 
So, the attacker's \inteighty injection introduces a new state $S_{a'}$, where $a'$ captures the result of executing the \inteighty handler.

\paragraph{Scope of Syscalls \& Registers.}
The attacker has a choice of invoking all syscalls by injecting \inteighty.
As shown in the \texttt{write} syscall example, the attacker needs to have precise arguments in general purpose registers: \eax should hold the correct syscall number and \ebx, \ecx, and \edx should hold the correct syscall arguments. 
Then, depending on register states, an attacker can change \eax and memory (arguments passed by reference) with syscalls. 
Identifying code locations in applications that satisfy this requirement, if not impossible, is challenging. 
To reduce the search space, we limit our analysis to syscalls that only depend on \eax. 
We analyze $328$ syscalls and find $40$ syscalls only take \eax as an argument and return \eax i.e., independent of other registers  (e.g., \texttt{getpid}, \texttt{getmask}, and other getter functions).
\texttt{sigreturn} uses the current user stack to restore the process stack and can be used for code reuse attacks.
Similarly, \texttt{setsid} creates a new session and process group and can be used to modify the value of \eax. 
Next, we assess which of these syscall invocations are of interest to an attacker. 
It is unlikely that at an interesting point during a program's execution \eax will hold the value of one of these syscalls.
\eax usually stores the return value of functions, so it often contains pointers and error values.
While we cannot meaningfully change pointer values by invoking syscalls, we observe that we can change returned error codes as shown in Sec.~\ref{sec:attack-ovw}.
However, it raises the question: is such a primitive too weak to bring about any malicious effects?

\paragraph{Altering \eax to non-zero value.}
Often guard conditions check for non-zero values, which if maliciously altered, can induce data and control flow changes, as shown in Rowhammer~\cite{rowhammer} and non-control-data attacks~\cite{non-control-data}. 
Thus, we make the conscious choice to restrict ourselves to only use the \inteighty gadget with \eax equal to zero (e.g.,  change the return value from 0 to -4).
Our case studies in Sec.~\ref{sec:case-studies-int80} show that this is a powerful primitive in itself.

\subsection{Signals to Userspace}
x86 architecture maps floating point exceptions (e.g., divide by zero, overflow) to interrupts. 
When these interrupts occur, the Linux kernel handles them and raises a signal (SIGFPE) to the user-space application. 
Applications can register user-space handlers for these signals which are executed when the kernel raises the signal.
We surveyed open-source applications that register explicit effect handlers for these signals. 

\paragraph{int 0, 9, and 16: Floating Point Exceptions (FPEs).}
We found that of all the signals that the kernel raises because of interrupts, SIGFPE is the most interesting. 
Handlers for SIGFPE perform operations like setting variables to certain values (e.g., set the denominator to a non-zero value to handle a divide-by-zero), or skipping some operations (e.g., ignore faulting data that cause overflows). 
Therefore, a malicious hypervisor can change the control and data flow of applications by triggering interrupts that raise SIGFPE. 
\begin{lstlisting}[language=C,escapechar=|]
/* Example: SIGFPE handling */
double arr[] = {...}
double weights[] = {...}
double avg = 0
void handler() { /* compute non-weighted avg */ } |\label{line:fpe:handler}| 
int compute_weighted() {
  register(SIGFPE, handler) |\label{line:fpe:register}| 
  avg = ...      /* compute weighted avg */ |\label{line:fpe:cause}| 
  ... |\label{line:fpe:inject}| 
  return avg
}    
\end{lstlisting}
For example, in the code snippet above, the application registers a SIGFPE handler on line~\ref{line:fpe:register}. 
If the computation on line~\ref{line:fpe:cause} causes a SIGFPE, the handler is executed. 
Then, the execution continues on line~\ref{line:fpe:handler}.
An attacker can inject the divide-by-zero interrupt on line~\ref{line:fpe:inject}.
This forces the application to always execute the handler changing its execution. 
As a result, the function always computes a non-weighted average compromising its integrity. 
Therefore, by injecting \intzero an attacker can introduce a new state $S_{a'}$ in the program's execution state (see ~\cref{sec:trace-reasoning}).

Note that, unlike the attack using \inteighty gadget which always invokes a syscall, the gadgets for FPE rely on application-specific handlers in user-space.
Further, if the application does not register a handler, the kernel uses a default handler that terminates the process.

\begin{figure}[t]
    \centering
     \includegraphics[scale=0.57]{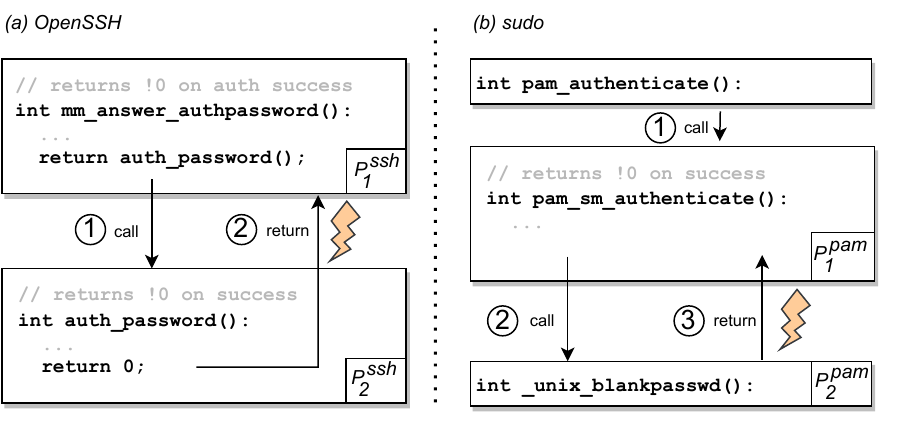} 
         \caption{(a) \ponesshd and \ptwosshd: gadget pages in the \openssh binary. (b) sudo \ponesudo and \ptwosudo: gadget pages in the pam shared library used by the sudo binary.}
    \label{fig:case-study-gadgets}
 \end{figure}

\paragraph{Other Signals.}
\codename can inject interrupts that generate  SIGTRAP (1), SIGILL (6), SIGSEGV (4, 5, 10), and SIGBUS (11, 12, 17, 29) signals to userspace applications.
However, we did not find applications that registered explicit effect handlers for these four signals.
In the absence of handlers, POSIX standard states that userspace application must be terminated. Thus, these four signals are uninteresting for \codename.

\paragraph{Chaining Interrupts.}
A malicious hypervisor can chain multiple gadgets by injecting interrupts at different points during an application's execution.  
For example, consider an application that performs multiple authentication checks and registers a SIGFPE handler. 
To successfully authenticate, the attacker should compromise the data flow on lines~\ref{line:chain:1} and~\ref{line:chain:2}.
First, the attacker uses \inteighty to bypass the check on line~\ref{line:chain:1}.
Then, after line \ref{line:chain:trigger:fpe}, the attacker triggers SIGFPE to change the value of \texttt{n} to $0$.
This changes the execution on line~\ref{line:chain:2} passing the second check. 
\begin{lstlisting}[language=C,escapechar=|]
/* Example: Chaining interrupts */
int n = 1
void handler() { n = 0 }
int auth() { return 0 }
void grant_access() {
  register(SIGFPE, handler)
  if (!auth()) { ... } /* deny access      */ |\label{line:chain:1}| 
  n = second_auth()    /* !0 if auth fails */ |\label{line:chain:trigger:fpe}| 
  if (!n) { ... }      /* auth. success    */    |\label{line:chain:2}|  
}        
\end{lstlisting}
\vspace{2pt}

\begin{figure}[t]
    \centering
     \includegraphics[scale=0.6]{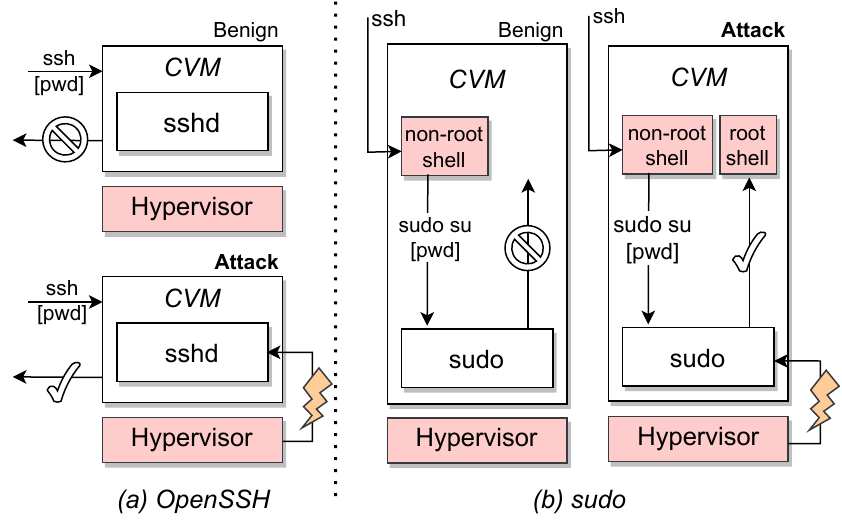} 
         \caption{Red: attacker-controlled, lightning: \inteighty injection, (a): Attack on OpenSSH, a malicious hypervisor successfully authenticates ssh on CVM with wrong pwd. (b): Attack on sudo, a malicious hypervisor with non-root shell on CVM escalates privilege to root shell.}
    \label{fig:case-studies}
 \end{figure}
\spacesave
\section{Case Studies}
\spacesave
We choose open-source applications to demonstrate the feasibility and impact of \codename. Then, we identify gadgets that allow a malicious hypervisor to mount \codename.
\label{sec:case-studies}
\label{ssec:openssh}

\subsection{int 0x80}
\label{sec:case-studies-int80}
\paragraph{\openssh.}
It allows authenticated users to obtain a secure shell, use subsystems (e.g., sftp) for file transfers, and execute commands on remote servers. 
In our threat model, bypassing \openssh's authentication imparts the attackers with powerful capabilities to compromise the execution of a CVM. 
To this end, we demonstrate an attack on an \openssh server on the CVM using  \inteighty as shown in~\cref{fig:case-studies}(a).
We assume a malicious hypervisor that does not have the correct root password to authenticate a secure-shell on the CVM. 
As shown in~\cref{fig:case-study-gadgets}(a), we identify a gadget where changing the return value to a non-zero number leads to successful authentication.
Specifically, our attack sets the return value of \authpasswd to a non-zero value using \inteighty.

\label{ssec:sudo}
\paragraph{Sudo.}
Using sudo, an authorized non-root user can escalate privileges to a root user. 
We demonstrate an attack on sudo where an adversary with access to a non-root shell on the CVM can gain root access (see~\cref{fig:case-studies}(b)). 
Specifically, the malicious hypervisor uses \inteighty to bypass sudo's authentication mechanisms. 
By default, sudo is configured to use Privileged Access Management (PAM).
With PAM enabled, sudo invokes a PAM module to authenticate the user.  
We identify a gadget in the PAM module with the \pamsmauth function as shown in~\cref{fig:case-study-gadgets}(b). 
This function first checks if the user has a blank password by calling the \blankpwd function. 
If this check succeeds, the PAM module does not prompt the user for a password. 
Instead, it considers the user to be correctly authenticated and returns to sudo.
Therefore, we can use \inteighty to change the return value of \blankpwd to a non-zero value leading to successful authentication. 
Applications that use the same PAM library to authenticate a user (e.g., doas~\cite{doas}) are also susceptible to \codename in principle.

\paragraph{Chaining \openssh and Sudo.}
\label{ssec:chaining}
\openssh can be configured to prevent login as the root user. 
Similarly, sudo can be configured (using the sudoers file) to limit the users who can execute it. 
With this setup, our attack using \openssh can only get a non-root shell and our attack using sudo is not possible. 
However, we can chain the two attacks to get past these issues. 
Specifically, we attack \openssh to get a non-root shell of a user in the sudoers list. 
This ensures that the non-root user can execute sudo. 
Then, we use the attack on sudo to escalate the non-root shell to root privilege as explained above.
Note that, to successfully chain the attacks, the malicious hypervisor injects \inteighty two times. 
\begin{figure}[t]
    \centering
     \includegraphics[scale=0.75]{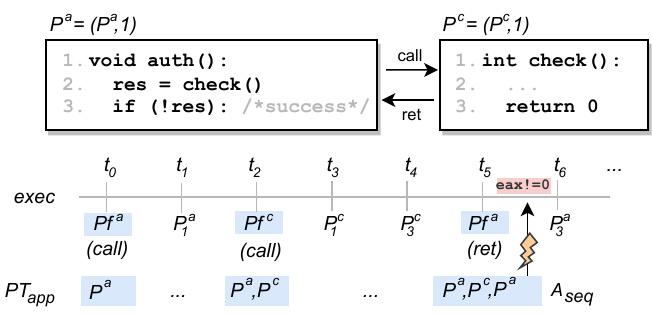} 
         \caption{Attacker bypasses authentication check by injecting interrupt at time $t_{5}$ when detecting \attackseq. Superscript for $P$: page id, subscript for $P$: line number in page, $Pf^{a}$: page fault in page with id $a$. For every page fault (blue), the GPA of the page is added to the \tapp. }
    \label{fig:timing}
 \end{figure}
\subsection{Applications with SIGFPE}
\label{ssec:apps-sigfpe}

We first surveyed language support for signal handlers and then looked for existing applications that register SIGFPE handlers with explicit effects.

\paragraph{Java Statistical Analysis Tool.} In Java, the runtime (Java virtual machine or JVM) registers a handler for SIGFPE in the user-space. 
When it receives SIGFPE from the kernel, the JVM translates it to a language-level ArithmeticException. 
The ArithmeticException is then caught and handled in the application. 
We analyze open-source Java applications that catch the ArithmeticException. 
We find an interesting gadget in the Java Statistical Analysis Tool (JSAT)~\cite{jsat-github}: a function that is used to add new data to a distribution that recalculates the mean and covariance as shown below. 
\\
\begin{lstlisting}[language=java,escapechar=|]
/* Example: Disrupt Java with SIGFPE */
try {
  Vec newMean = ...;        /* new mean */ |\label{line:java:inj}| 
  Matrix covariance = ...;  /* new covariance */
  this.mean = newMean;
  setCovariance(covariance);
} catch(ArithmeticException ex)
{ this.mean = origMean; }
\end{lstlisting}
During normal execution, if the function catches an ArithmeticException it uses the original mean, effectively ignoring the faulting data. 
On line~\ref{line:java:inj}, a malicious hypervisor can inject an interrupt that raises SIGFPE (e.g., \intzero for divide-by-zero) and consequently the ArithmeticException to the application. 
This will ensure that the function always ignores any new data added. 
This gadget is used to add new data to a multivariate normal distribution. 
Therefore, our attack can be used to bias the distribution to never accept new data. 

\paragraph{TextAnalysis.jl in Julia.} 
Like Java, the Julia runtime forwards signals for SIGFPE to a language-level DivideError. 
We find an interesting gadget in an established Julia package for text analysis (TextAnalysis.jl)~\cite{julia-textanalysis-github}: an evaluation function to calculate a performance metric based on precision and recall scores (F-Score). If the function catches a DivideError, it reports the worst performance, indicating that a pair of text (e.g. machine and human-produced) are not similar.
\begin{lstlisting}[language=Julia,escapechar=|]
# Example: Disrupt Julia with SIGFPE
function fmeasure_lcs(RLCS, PLCS, beta=1)
  try
    return ((1+beta^2) * RLCS * PLCS) /
           (RLCS + (beta^2) * PLCS)
  catch ex
    if ex isa DivideError
      return 0
    ...  
\end{lstlisting}
We leverage this by maliciously raising SIGFPE and consequently DivideError to report the worst performance.

\paragraph{Hand-coded Multi-layer Perceptron (MLP) in C.}
We take an MLP implementation written in C~\cite{mlp-github} that uses \texttt{tanh} from the math library as an activation function as shown in the code snippet below. 
We manually add a SIGFPE handler, that recovers from overflows by setting the return value to $1$ as shown in the code snippet below. 
\begin{lstlisting}[language=C,escapechar=|]
/* Example: Disrupt MLP with SIGFPE */
void tan_h_classify(...) {
  output[0] = 1            /* bias term */
  for (i = 0; i < n; i++)
    if (sigsetjmp(buf, 1)) /* on SIGFPE */
      output[i+1] = 1 
    else                   /* no overflow */
      output[i+1] = tanh(input[i]) |\label{line:mlp:inj}| 
}
\end{lstlisting}
We then maliciously invoke the handler to bias the model trained by the MLP. 
Specifically, on every call to the \texttt{tanh} function, we inject the interrupt to trigger SIGFPE (line~\ref{line:mlp:inj} in the code snippet below). This ensures that the \texttt{tanh} function always returns 1. 
This allows us to bias the final confusion matrix for our test data set.

\spacesave
\section{When \& Where to Inject Interrupts?}
\label{sec:timing-injection}
\begin{figure}[t]
    \centering
     \includegraphics[scale=0.65]{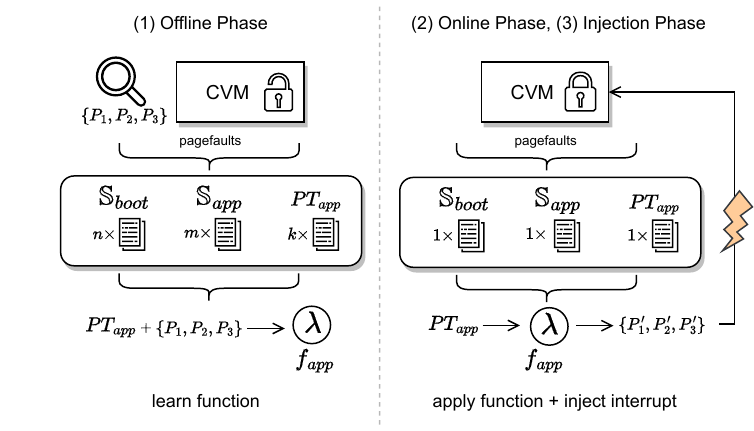} 
         \caption{Overview of profiling. During offline phase (1) we learn a function (\fapp) that maps pagefault patterns to \codename gadgets. We repeatedly create \sboot, \sapp and \tapp. During online phase (2), we apply the function to monitor when the CVM reaches a point of interest in its execution, in the injection phase (3) we inject the interrupt. $\{P_1, P_2, P_3\}$: physical addresses of \codename gadget pages in \tapp, $\{P'_1, P'_2, P'_3\}$: predicted \codename gadget pages in \tapp.}
    \label{fig:overview}
 \end{figure}

For our attacks to succeed, it is crucial that we inject the interrupts at specific points during the application's execution.
For example, to attack \openssh (see~\cref{ssec:openssh}) we should inject the interrupt before the \mmanspwd uses the value returned by \authpasswd as shown in~\cref{fig:case-study-gadgets}(a). 
If we inject the interrupt at other points during the application's execution, the injection might not have the desired side-effect (e.g., changing \eax before it is used), or crash the application. 
Next, if the CVM has multiple VM cores, we should ensure that our interrupt injection is targeted to the right core that executes the application logic with our gadget. 

\paragraph{Overview.} For SEV-SNP, the main challenge for a successful attack is identifying the physical pages of the functions of interest (i.e., \mmanspwd and \authpasswd for \openssh).
By marking the stage-2 page tables as non-executable we can trace the transition from \authpasswd to \mmanspwd.
This is possible because our two target functions are on two different physical pages. If this is not the case, i.e., both the functions are on the same page, we will have to resort to single-stepping this part of the execution~\cite{sev-step}. 
However, for our builds of the target libraries, the functions are indeed on different pages. 
Therefore, once we observe a page fault on \authpasswd followed by a page fault on \mmanspwd, we inject \inteighty.
Specifically, every stage-2 page fault causes a VMEXIT transparent to the CVM. This allows \codename to inject an interrupt when the VM resumes, right before it executes the next guest instruction. In summary, the VM uses the attacker altered state on resumption from the page fault.

\paragraph{Attack Phases.}
\codename attack requires three phases: (a) an offline analysis to learn a function (\fapp) that maps page fault patterns to \codename gadgets; (b) an online analysis to monitor when the CVM reaches a point of interest in its execution; and (c) injecting the interrupt (see~\cref{fig:overview}). 
In the offline phase, we assume that the malicious hypervisor can create and run CVMs identical to the victim CVM multiple times to profile the behavior of the victim applications~\cite{cachewarp}.
In this phase, the attacker controls both the malicious hypervisor and the CVM. 
In the online phase, when the attacker injects the interrupt, the attacker only controls the malicious hypervisor but can observe the CVM. 
Next, we detail how \codename uses the different phases to learn the \fapp function.

\paragraph{Page Traces.}
To time and target our interrupts to the right core, we rely on the fact that our gadgets sequentially execute functions on different pages as shown in~\cref{fig:timing,fig:case-study-gadgets}.
Let us assume that the hypervisor can capture all pages with the cores they were used on during a CVM's execution (e.g., using page faults). 
Specifically, the hypervisor captures a list ($PT_{\tt{vm}}$) of tuples with the guest physical addresses (GPAs) of pages and their corresponding cores $[(p^{\tt{id}}, \tt{core})]$. 
Using \ptvm, the hypervisor creates application-specific \tapp shown in~\cref{fig:timing} with all pages executed by the app in user-space.

The code snippets in~\cref{fig:timing} are analogous to the gadgets we detail for our case studies in~\cref{sec:case-studies}. 
Here, the \auth function on page \pagea calls \checkf on \pagec and uses its return value. 
Therefore, the application's page trace (\tapp) always contains the sequence \attackseq $=$ [\pagea, \pagec, \pagea]. 
To time the interrupt and target the right core, the hypervisor observes the application's access to these pages and waits to detect the sequence of pages. 
When the hypervisor detects the sequence \attackseq it injects the interrupt (e.g., \inteighty to change the return value of \checkf) before execution resumes on line $3$ on \pagea ($P^{a}_{3}$ in~\cref{fig:timing}).
Note that \tapp is sufficient to target the interrupt to the right core as it contains information about the core on which the page was accessed by the application.

\paragraph{Application Trace (\tapp).}
To capture \ptvm, we assume that the hypervisor can induce page faults for all page accesses in the CVM. 
Creating \tapp from \ptvm is not straightforward. 
First, the GPAs for the application's pages are different for every execution. 
Next, \ptvm contains pages used by the kernel and all user-space applications. 
Further, the order in which the pages are accessed in the CVM is affected by the scheduling decisions in the Linux kernel. 
Given these challenges, we detail a method to reliably create \tapp and identify \attackseq. 

Capturing every page access for a CVM's execution (\ptvm) is expensive (many page faults for the same page) and generates an intractable trace. 
Instead, it is sufficient to start with a set of pages executed when the victim application executes on CVM (\svm). Note that, this only requires $1$ page fault per page that is executed on the CVM. 
\svm contains some pages executed by the kernel that need to be removed while creating \tapp. 
To identify the kernel's pages, we capture the set of pages accessed during kernel boot to form \sboot. 
By removing all pages in \sboot from \svm we get \suser i.e., \suser $=$ \svm $\backslash$ \sboot. 
Now, \suser contains all user-space pages executed in the CVM. 
To eliminate pages that do not belong to our victim application (e.g., \openssh, sudo) we execute the application multiple times ($n$) and compute \susern for every iteration ($i$). 
The set intersection of all \susern gives us \sapp i.e., \sapp = $\bigcap_{i=1}^{n}\mathbb{S}_{\tt{user}_{\tt{i}}}$. 
By increasing the value of $n$, we can ensure that \sapp only contains pages executed by our application. 

Once we have correctly identified the application's pages, we can capture the pages in \sapp every time they are executed to form \tapp. 
The guest physical addresses of the application's pages change when the VM is rebooted. 
Therefore, to reliably find our gadget pages (\pagea and \pagec) we should account for the changing GPAs. 
To capture this, we collect \tapp over multiple VM boots. 
Then, we analyze all \tapp to find a function \fapp to get the gadget pages \pagea and \pagec in~\cref{fig:timing}.
Finally, we can use the gadget pages to identify \attackseq to correctly time and target the interrupt injection.

\spacesave
\section{Implementation for AMD SEV-SNP}
\label{sec:impl}

We describe our method to identify the guest physical address of the page that houses the gadgets of our interest.

\subsection{Generating Page Traces}
\label{ssec:gen-page-traces}
To generate the page trace for the application (\tapp), we need to induce page faults every time a page in the CVM is executed. 
In SEV-SNP the hypervisor can force page faults in the CVM~\cite{sev-step, SEVered}. 
\sevstep implements a mechanism that can be configured to induce page faults on all pages, or only on $1$ page. 
We use the former configuration to create the unordered sets described in~\cref{sec:timing-injection}. 
Specifically, before booting the VM, we mark all pages as not-executable by setting the \nx bit. 
Every time a page fault occurs, we note the page's GPA and core. 
Before the CVM resumes execution, KVM clears the \nx bit. 
This ensures that only $1$ page fault is triggered per page. 
To create the ordered list (\tapp) we use the mechanism from \sevstep to mark single pages as not-executable. 
We start by setting all pages in \sapp as not executable. 
Then, on every page fault, we note the GPA and core. 
Next, we set the \nx bit of the page that generated the previous page fault. 
This mechanism ensures that every access to the application's pages generates a fault. 

To implement this mechanism, we use the modified KVM from \sevstep which exposes ioctls to the user-space~\cite{sev-step}. 
These ioctls allow user-space applications to register and wait for events (e.g., page faults). 
We create CPython (409~\loc) and Python programs (2291~\loc) to interface with KVM to register and handle events for page faults. 

\paragraph{Optimization.} If we enable page faults for all application pages, the size of \tapp grows. 
We know that our gadget pages will only be accessed a few times during the application's execution. 
Therefore, we define an upper limit on the number of occurrences of a particular page in our tracing. 
This reduces \tapp size and optimizes the application execution time. 
\spacesave
\subsection{Boot Set (\sboot) and Application Set (\sapp)}
\label{subsubsec:sapp}
In both the offline and online phases of the attack, to create the application page trace (\tapp), we first need to form the boot set and application sets for each case study.

\paragraph{Boot set.} We use the boot set to eliminate all pages executed by the kernel from \tapp. 
To create this set, we mark all pages as not-executable before booting the CVM. 
We capture all pages that generate page-faults while the Linux kernel boots on the CVM and add them to the boot set.  
We stop the capture once the CVM boot completes.  
This ensures that only kernel pages are captured in the boot set. 
Next, we explain how we create the application set for our end-to-end case studies.

\paragraph{\openssh.}
For password authentication, \openssh prompts the user for a password. 
If the authentication fails, it prompts the user again. 
The code gadgets we are interested in (see~\cref{sec:case-studies}) are executed between these successive prompts. 
Therefore, to form the application set for \openssh, it is sufficient to capture the pages that are executed in this password prompt window. 
To do this, we implement a Go program as an ssh client with $70$~\loc. 
For fine-grained control over the password authentication process, we modify Go's \texttt{crypto/ssh} standard library.  
We execute the ssh client from the untrusted host multiple times and capture the pages that are executed to form \susern and subsequently \sapp as explained in~\cref{sec:timing-injection}. 

\paragraph{Sudo.}
It uses PAM to perform password authentication by calling the \texttt{pam\_unix} shared library which has our code gadget from ~\cref{sec:case-studies}. 
The Linux kernel executes shared libraries from the same physical addresses. 
Thus, for all executions of the shared library, the GPAs remain constant.
We use this fact to create our application set for the sudo binary. 
Specifically, we write a C program 
to repeatedly access the pages with our code gadget i.e., \ponepam and \ptwopam of the \texttt{pam\_unix} library shown in~\cref{fig:case-study-gadgets}(b) as shown below.
\begin{lstlisting}[language=c,label={lst:catchguestpa}]
/* Profiling shared libraries */
char* lib = "/usr/lib64/security/pam_unix.so";
unsigned long gad1, gad2; char* a; int fd; 
fd = open(lib, O_RDONLY);
a  = mmap(0, 0x4000000, (PROT_READ | PROT_EXEC), 
          MAP_SHARED, fd, 0);               
gad1 = a + 0xCAFEBABE; /* ret gadget 1 */
gad2 = a + 0xCAFED00D; /* ret gadget 2 */
while (1) {
  asm volatile("mfence":: :"memory");
  asm volatile("push %
  asm volatile("jmp *%
jmp1:
  asm volatile("mfence":: :"memory");
  asm volatile("push %
  asm volatile("jmp *%
jmp2:
}
\end{lstlisting}
We execute this C program several times on the CVM and capture the pages that are executed to create \sapp.
Note that, the addresses for \texttt{pam\_unix} are fixed, so we do not need to execute sudo application during this phase to form \sapp.

\paragraph{MLP.}  
We use an open-source implementation of MLP written in C~\cite{mlp-github} and add a SIGFPE handler to its tanh activation function implementation. 
Every call to this activation function results in multiple calls to the \texttt{tanh} function in the math shared library as shown in~\cref{ssec:apps-sigfpe}.
We implement an interface in the CVM to allow users to start and stop the training of the MLP. 
The training process takes a long time. 
Therefore, capturing all pages executed during the training results in a very large unusable set. 
So, we capture the pages multiple times during the training in small windows of 1 second.
To form \susern we compute an intersection over all pages from the windows to create \sapp(see~\cref{sec:timing-injection}).
\spacesave
\subsection{Finding a Function}
\label{ssec:finding-function}
In the offline phase, we use the application page trace (\tapp) to define a function (\fapp) to predict the physical addresses of our gadget pages. Next, we explain our how to create \fapp for each of the end-to-end case studies.  

\paragraph{\openssh.}
We analyze the page traces (\tapp) from multiple CVM boots. 
We first create 2 sets with potential candidates for gadget pages (\ponesshd and \ptwosshd). 
Using the page traces across multiple CVM boots we profile the \openssh behavior during password authentication and define a frequency interval $[9,11]$.
We define all pages that appear in \tapp with frequencies in this interval as candidate pages for \ponesshd. 
Similarly, we define a frequency interval $[5,7]$ to find the candidate pages for \ptwosshd. 
Note that, for these VM boots, the attacker also controls the CVM. 
During the attack, we first form the candidate sets using the values for the frequency intervals we define above. 
Then, to further eliminate pages from the candidate sets and form page tuples (\ponesshd, \ptwosshd), we use the fact that the gadget pages must appear in a particular sequence (\attackseq) in all page traces. 

\paragraph{Sudo.} 
Unlike \openssh identifying the gadget pages is straightforward for sudo. 
First, in this setting the attacker already controls a non-root shell on the CVM. 
Then, our gadget pages lie in the \texttt{pam\_unix} shared library whose GPAs do not change across multiple runs.  
The C program's loop uses the virtual addresses of the gadget pages to repeatedly access them (see~\cref{subsubsec:sapp}). 
To determine the GPAs of these gadget pages our function (\fsudo) just picks the $2$ pages that occur the most number of times in \tapp. 
Then, it uses the order of accesses to determine the GPAs for the tuple (\ponepam,\ptwopam).

\paragraph{MLP.}
We identify $3$ gadget pages for MLP: \ponemlp the page that contains the calling function of tanh, \ptwomlp the tanh shared library, and \pthreemlp a page in the shared library executed by the tanh function. 
While the first page is backed by different GPAs for each application execution, the second and third page in the shared library remain constant. 
On investigating the application trace \tapp, we identify a sequence of length $9$ with the gadget pages that occur with high frequency. 
We use this to define the function (\fapp) the finds candidates for tuples of gadget pages (\ponemlp, \ptwomlp, \pthreemlp).

\paragraph{Effect of Imperfect Page Analysis.}
Our AMD SEV-SNP analysis is intentionally specific to our observations per application. It is not designed for other gadgets that may not conform to such behaviour, and depending on the gadget, may need instruction single-stepping~\cite{sev-step, cachewarp}.
Injecting \inteighty on the wrong page either has no observable effect or crashes OpenSSH which is restarted by the daemon.

\paragraph{Remark on Intel TDX.} 
We need a primitive to know when the \attackseq occurs during the application's execution.
Since our goal is not to build single-stepping and analysis techniques demonstrated for AMD SEV~\cite{sev-step}, we do not investigate using page faults, cache side-channels, or timer interrupts, to achieve this primitive.
We had limited access to the TDX machine to fully experiment.
To make the best use of our limited access and to demonstrate our attack, we use a busy loop in functions \mmanspwd and \pamsmauth for \openssh and sudo respectively.
Future works can address this using advances in TDX-step~\cite{tdx-step}.

\section{Proof-of-concept Exploits}
\label{sec:e2e-attacks}

\label{ssec:setup}
To demonstrate \codename on SEV-SNP and Intel TDX, we use the latest production systems and setups recommended by AMD and Intel respectively.

\paragraph{SEV-SNP.}
We demonstrate our attacks on an EPYC 9124 with Zen 4 SEV-SNP enabled workstation with 16 cores and 192\,GB RAM. 
We boot the host Linux kernel with patches from \sevstep that introduce the page-fault interfaces in KVM~\cite{sev-step}. 
This kernel also contains the patches for KVM to launch and manage SEV-SNP VMs. 
Further, we use the same QEMU version 6.1.50 and SEV-SNP VM Linux kernel v5.19.0  to perform our experiments.

\paragraph{TDX.}
We had early access to TDX in September 2023.
We confirm our attacks on a pre-production Intel Xeon Platinum processor with TDX support with 112 cores and 256GiB of RAM. 
We follow the official Intel documentation and boot a patched Linux kernel v5.19.17 on both the guest and the host. 
Further, we use modified QEMU v7.0.50 provided by Intel to create TDX VMs. 
In March 2024, we tested \inteighty injection on a production Intel Xeon Gold 6526Y processor with TDX support and confirmed that it is vulnerable to \codename.

\spacesave
\subsection{Injecting Interrupts}
\label{sec:inject-interrupts}
While both SEV-SNP and TDX allow the hypervisor to inject interrupts to the CVMs, the method to inject the interrupt is different for each of them. Below, we outline the mechanisms we use to inject interrupts for \codename.

\paragraph{SEV-SNP.} %
AMD virtual machine extensions expose various interfaces that a hypervisor can use to inject interrupts into a VM.
In our implementation, we use the event injection interface to inject \inteighty and \intzero (see \cref{appendix:injection} for other interfaces).
For this, we use the event injection field (\texttt{VMCB.EventInj}) that is accessible to the hypervisor in the Virtual Machine Control Block (VMCB) of the SEV VM. 
We implement a kernel module with 150 \loc, which interfaces with KVM to write the interrupt number to be injected in the respective VMCB field. 
When the SEV VM resumes execution, this method ensures that the interrupt is always raised before the next instruction is executed~\cite{amd-manual}.
This makes our injection deterministic, thus ensuring \codename does not need to time the interrupt injection between a window of a few CPU cycles as already explained in~\cref{sec:timing-injection}.
The KVM implementation expects acknowledgments from the guest kernel in the VM for most external interrupts. 
During normal operation, for all external interrupts, the guest Linux kernel writes the acknowledgments to a register in this virtual APIC page. 
We observe that the \inteighty handler in the guest Linux kernel does not acknowledge the interrupt because it does not expect these interrupts to be injected externally. 
Without such acknowledgment, KVM will not inject certain interrupts which can lead to unexpected behavior (e.g., frozen terminal because of tty interrupts).
To remedy this, we perform the virtual APIC page register write from the host.

\paragraph{TDX.}
We implement a kernel module in 150 \loc to inject interrupts into the TDX VM. 
Our host module uses kernel hooks to call a function in KVM that is used to deliver \inteighty interrupts to TDX VMs.
Unlike SEV-SNP, TDX does not expose the Virtual Machine Control Structure (VMCS) or the virtual APIC pages to the untrusted hypervisor. 
Instead, it expects the hypervisor to write into a Posted Interrupt Request (PIR) buffer. 
This buffer is used by hardware to inject interrupts into TDX VMs through the virtual APIC~\cite{intel-sdm}. 
We inject two interrupts into two different cores of the CVM with this mechanism, one to gain login into the TDX VM with \openssh and another to get root access with sudo.
During these two injects, the guest kernel does not acknowledge the interrupts. 
While this does not stop our attacks, it does leave the APIC with an elevated Task-Priority-Register (TPR), blocking all lower-priority interrupts on the affected vCPU.
This may break CVM functionality that is noticeable by the user. 
To evade such detection, we implement a guest kernel module (\kernack) that resets the APIC state. 
We inject this kernel module into the TDX VM as the last part of our attack after gaining root access.

\spacesave
\subsection{\openssh}
\begin{table}[]
\renewcommand{\arraystretch}{1.2} 
    \caption{Cardinality of the sets (\sboot, \suser, \sapp) and traces (\tapp) to find gadget pages. VMb: VM boot, max captures: maximum number of times we capture a page in \tapp where 0 indicates that we always capture.}
    \spacingtable
    \label{tab:traces}
    \resizebox{\columnwidth}{!}{%
    \begin{tabular}{@{}lrrrrrrr@{}}
    \toprule
            & VMb & \begin{tabular}[c]{@{}l@{}}traces\\ per VMb\end{tabular} & \sbootcardinality & \susercardinality & \sappcardinality & \tappcardinality &\begin{tabular}[c]{@{}l@{}}max\\ captures\end{tabular}  \\ \midrule
    Openssh & 392   & 10    & 82433 &  666 & 236 & 22440 & 200 \\
    Sudo     & 9 & 1    & 82431 & 259   & 6 &   199013 & 0 \\ 
    MLP   & 6 & 20  &  82378 & 718 &  255 &  32832 & 200  \\
    
    \bottomrule
    \end{tabular}%
    }
    \end{table}

We do our attack on an \openssh binary v9.4.P1+ with PAM disabled. We run an ssh client on the same host as the CVM. 

\paragraph{SEV-SNP.}
In the offline phase, we profile the behavior of \openssh over 392 VM boots. 
For every VM boot, we collect 10 user sets (\suser).
Using these, we create the application set (\sapp) and page traces (\tapp) of sizes shown in~\cref{tab:traces}.

In the online phase, to profile and attack the application, we set up the VM to generate page faults during boot and during application execution. 
With the page fault mechanism enabled, we observe an overhead of 11.41 seconds to boot as compared to 10.01 seconds without the page faults (+14\%). 
Creating the application set in the password prompt window takes 32.9~ms to execute compared to 14.2~ms without page faults (+131\%). 
Creating a page trace \tapp for the password prompt window takes 773.9 ms to execute (+5332\%). 
In \Cref{tab:trace:overhead}, we summarize page fault overheads for all the case studies.
For \openssh and MLP, we cap the number of times each page is captured to 200 (see~\cref{ssec:gen-page-traces}).
\Cref{tab:gadget-pages} shows the number of times our gadget pages appear on average in \tapp.
From our profiling, we report that on average the size of our candidate set for \ponesshd is $4.81$, and \ptwosshd is $8.52$ before considering the attack sequence. 
Finally, when we account for the sequence (\attackseq) in \tapp, on average we get $2.24$ (\ponesshd, \ptwosshd) tuples.
With this, we get an average probability of success of $44.71\%$ with $1$ interrupt injection.

\paragraph{TDX.}
As explained in~\cref{sec:impl}, we implement busy loops in our gadget page with the function \mmanspwd. This eliminates the need to time our interrupt injection. 
We use our kernel module in the host to inject \inteighty. The interrupt-delivering function takes 1835 cycles for every injection. 
Further, once the attack succeeds, we insert a kernel module in the TDX VM to reset the APIC. 
The reset takes about $3092$ cycles on average. 
With this setup, we report that our attack always succeeds.

\begin{table}[]
\caption{Number of times (in \% and absolute) the gadget pages for the different applications appear in the application's page trace (\tapp). Page trace size (\tappcardinality) as detailed in \Cref{tab:traces}. The gadget page $P_{3}$ is not applicable to \openssh and sudo as they only have 2 gadget pages. }
\spacingtable
\centering
\renewcommand{\arraystretch}{1.1} 
\small
\label{tab:gadget-pages}

\begin{tabular}{rrrrrrr}
\toprule
\textbf{}   & \multicolumn{2}{c}{\openssh} & \multicolumn{2}{c}{Sudo} & \multicolumn{2}{c}{MLP} \\

\textbf{}   &  \multicolumn{1}{c}{\%}      & \textit{abs.}   & \multicolumn{1}{c}{\%}  &\textit{abs.}  & \multicolumn{1}{c}{\%}  &\textit{abs.}    \\
\midrule
$P_{1}$ &      $0.044$  & $9.8$   &  $25.3$   &   $50348.6$   &   $0.6$  &  $200$               \\

$P_{2}$ &   $0.026$    & $5.9$   &  $24.6$   &  $49051.0$ &    $0.6$   & $200$                \\

$P_{3}$ &       \multicolumn{1}{c}{-}      &    \multicolumn{1}{c}{-}  &     \multicolumn{1}{c}{-}          &       \multicolumn{1}{c}{-}           &      $0.4$           &   $133$             
\\ 
 \hline
\end{tabular}
\end{table}

\begin{table}[]
\caption{Overheads for boot trace, application set (\sapp), and page trace (\tapp) w.r.t. execution without page faults in \%. }
\spacingtable
\label{tab:trace:overhead}
\centering
\small
\renewcommand{\arraystretch}{1.2} 
\begin{tabular}{@{}llll@{}}
\toprule
App     & boot & \sapp & \tapp \\ \midrule
OpenSSH & 14   & 131  & 5332  \\
Sudo    & 37   & 3    & 602   \\
MLP     & 38   & 3    & 81    \\ \bottomrule
\end{tabular}
\end{table}

\subsection{Sudo}
We use an unmodified sudo binary in the Ubuntu 23.10 distribution with default configurations. 

\paragraph{SEV-SNP.}
We perform our offline profiling over 9 VM boots and create the application set (\sapp).
To create an application trace (\tapp), we execute the loop that repeatedly accesses the shared library pages as explained in~\cref{subsubsec:sapp}. 
With this, we see that our gadget pages are in \sapp and up to $49.9\%$ of the final trace (\tapp) as shown in~\cref{tab:gadget-pages}.
For the attack, we execute \verb|sudo su| from the non-root shell on the CVM. 
Our looping technique to access the pages of the \texttt{pam\_unix} shared library ensures that we reliably find the GPAs of the gadget pages and our attack always succeeds with 1 injection.

\paragraph{TDX.}
To perform the sudo attack, we implement a busy-loop in the \pamsmauth function that waits for \inteighty. 
Therefore, our attack always succeeds and we escalate to a root shell on the TDX VM. 
To acknowledge the interrupt, we insert the kernel module as with the \openssh attack.

\paragraph{Chaining \openssh and Sudo.}
We chain our attacks on \openssh and sudo to get around the problems discussed in~\cref{ssec:chaining} by injecting \inteighty two times. 

\spacesave
\subsection{FPE}
We use three different applications to demonstrate \codename with interrupts that raise SIGFPE.

\paragraph{MLP.} 
To profile the MLP application offline, we record all pages that are executed in one second windows. 
We capture the pages over 6 VM boots and collect 10 user sets  (\suser) per boot. 
We observe average application set (\sapp) sizes of 255 pages. We create 20 traces (\tapp) per VM boot.
Our gadget pages $P_{1}$, $P_{2}$, $P_{3}$ occur $1.6\%$ in \tapp. 
Using our function from~\cref{ssec:finding-function} on average we find $20.5$ tuples for the gadget pages. 
We see an average probability of success of $41.6\%$.

\paragraph{JSAT and TextAnalysis.jl.}
Our method in \cref{sec:impl} requires more engineering to Java and Julia applications with runtimes (e.g., OpenJDK and Julia Runtime).
As opposed to ahead-of-time compiled programs, finding the gadget pages for interpreted programs requires profiling the dynamic behavior of the runtime's code cache and hot paths.
For simplicity, we run our programs with a busy loop in the gadget function instead of profiling it. We run the JVM in interpreter mode where SIGFPE is translated to a language-level ArithmeticException. 

For JSAT, we run the \texttt{LVQLLC} test to create a multivariate normal distribution from the JSAT repository~\cite{jsat-github}. 
With our attack, we need to inject $240$ interrupts while the application executes to change all return values of our gadget function (\cref{ssec:apps-sigfpe}).
Similarly, for TextAnalysis.jl, we run the \texttt{Evaluation Metrics} test suite from the TextAnalysis.jl repository~\cite{julia-textanalysis-github} and need to inject $2$ interrupts.

\subsection{End-to-End Attack Cost}
\codename is performed in $3$ different phases as shown in~\cref{fig:overview}. To understand the end-to-end cost of our attack, we explain the overheads for each of these phases.

\paragraph{Offline Phase.} During the offline phase, we get multiple traces as summarized in~\Cref{tab:traces}. In this case, the overheads of tracing slowdown the function generation described in Sec.~\ref{sec:inject-interrupts}. While this can be further optimized, we did not put efforts in such optimizations since this is a preparatory step before the victim runs its VM.

\paragraph{Online Phase.} \codename also enables page fault tracing in the online phase, i.e.~when the victim starts interacting with the VM.
Tab.~\ref{tab:trace:overhead} shows a timing analysis to generate boot trace, application set (\sapp), and page trace (\tapp) during online phase, when compared to the execution of the CVM without page fault tracing.
\codename causes some slowdown but it does not impact the victim's usability or result in detection. This is because we can potentially perform the tracing and the injection after attestation, but during the CVM provisioning which can take several minutes even in a benign setting. Thus, \codename attack happens before the user gets access to the VM, so it will not notice the lag. In cases where this is not possible, we can further cap the number of page faults for any given page (max capture in~\Cref{tab:traces}) to reduce the lag, as well as repeat the set intersection for \sapp to decrease the size of the resulting trace (\tapp). 

\paragraph{Interrupt Injection.} As already layed out in Sec.~\ref{sec:timing-injection} and \ref{sec:inject-interrupts}, we use SEV's event injection interface (\texttt{VMCB.EventInj}) to inject interrupts. This method ensures that the hardware raises the interrupt to the guest kernel before the VM executes subsequent instructions.

\spacesave
\section{Ineffectiveness of Current Defenses on AMD}
AMD SEV-SNP outlines two optional modes called Restricted and Alternate injections. 
They are designed to restrict the hypervisor's interrupt and exception interface to the CVM.
We explain the changes brought by these modes and then analyze their effectiveness against \codename. 

\begin{figure}[t]
    \centering
     \includegraphics[scale=0.7]{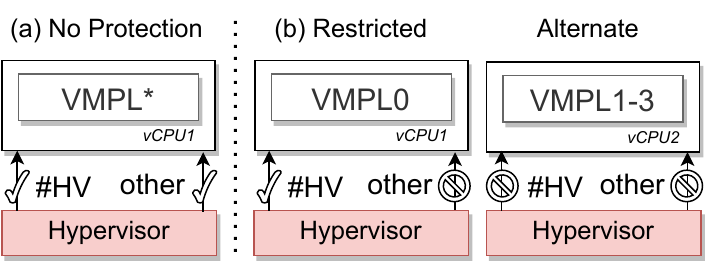} 
         \caption{(a): Without any defense enabled, the hypervisor can inject all interrupts into the SEV VM. (b) Restricted mode: enabled on vCPU that runs VMPL0 and the hypervisor can only inject \#HV. Alternate mode: enabled on all non-VMPL0 cores and the hypervisor cannot inject any interrupts.}
    \label{fig:defense}
 \end{figure}

\paragraph{AMD SEV-SNP Restricted Injection.}
The hypervisor sets bit 3 in the \texttt{SEV\_FEATURES} register per vCPU of the CVM to enable or disable this mode. 
When disabled, the hypervisor continues to use the legacy interfaces to inject {\em all} interrupts. 
When enabled, the hypervisor is still able to partially use the legacy interface 
(see \cref{fig:defense}(b)). 
Specifically, it can inject {\em only} \#HV interrupt---a new interrupt with number $28$ introduced for this mode. 
Further, the hypervisor cannot use the virtual interrupt queuing. 
Instead, the hypervisor and the CVM setup a shared memory region to house the event queue. 
The hypervisor uses the \#HV as a doorbell to inform the CVM about a new interrupt in the queue.
The \#HV handler in the CVM then accesses the queue, retrieves the actual interrupt number (e.g., \inteighty) and then handles the queued-up interrupt.

\paragraph{AMD SEV-SNP Alternate Injection.}
The restricted mode described above introduces a new interface for the hypervisor. More importantly, it breaks compatibility with existing guest OS implementations, requires enlightening the guest OS, and hinders lift-and-shift.
To limit this effect, the alternate injection mode offers the traditional interrupt interface, but with a caveat.
First,  one of the vCPUs in the CVM runs at a special privileged level called VMPL0 while the rest of the vCPUs execute at non-privileged levels VMPL1-VMPL3.
Second, all the vCPUs that execute the guest OS run in VMPL1-3 and enable alternate mode. With this combination, they continue to see a traditional interrupt interface both for configuring and receiving interrupts. 
Third, the vCPU that executes in VMPL0 acts as a trusted bridge between VMPL1-3 CPUs and the hypervisor. It also performs security and virtualization tasks within the CVM.
Since this is a new piece of code that is introduced, it can very well be in charge of presenting legacy interrupt interfaces for the CVM. This is why, it runs in restricted mode, creates a shared page, handles \#HV, converts them to virtual interrupts, and delivers them to the guest OS.
\cref{fig:defense}(b) shows the setup where both the modes are enabled on CVM cores. 
Note that both of these modes change the delivery mechanism and interfaces that the hypervisor needs to use to deliver the interrupts to the guest OS, it does not fundamentally introduce any filtering or dropping rules. 

\paragraph{Hardware Availability.}
Our machine supports both of these modes in the hardware and we were able to test that the newly introduced MSRs are operational.

\paragraph{Impact on \codename.}
The main goal of these new modes is to allow the CVMs to continue with their assumed behavior about the interrupt interface provided by the hypervisor for compatibility. The AMD documentation alludes that this mode can address potential misbehavior by the hypervisor that breaks the OS assumptions (e.g., inject interrupts while TPR is elevated). But, it does not discuss any mandatory security checks or filtering rules. 
The pseudo-code provided by AMD does not do any security checks.
More importantly, the software support for restricted mode does not perform any checks or filters~\cite{svsm-github}. 
Thus, even with restricted mode and \#HV, \codename attacks are possible. The main reason is that the new mode changes the delivery mechanism but does not stop or filter the delivery of interrupts. 
As for alternate mode, the current hypervisor and guest OS implementations do not support alternate mode. When implemented, it remains to see if it filters any interrupts, even though such filtering is not specified by AMD.

\spacesave
\section{Potential Defenses}
Given that existing mechanisms for interrupt security are insufficient, we develop software methods (where possible) and propose hardware mechanisms to mitigate  \codename. 
\spacesave
\subsection{Software Mitigations}
\label{sw:mitigations}
\paraspacesave
The main ingredient for \codename is the ability of the hypervisor to {\em externally} inject \spurious interrupts into a vCPU executing the CVM. 

\paragraph{Detecting External Interrupts.}
One seemingly straightforward fix is to address the symptom of external interrupts 
in software. For example, the guest kernel can be patched to detect and selectively allow external interrupts. 
Interrupts such as \inteighty, should perhaps never arise externally and can be dropped.
However, we did find use-cases where this is a desired behavior~\cite{xen-int80, inria-int80}, after all, it is part of the x86-64 ISA standard.
To disable external delivery of \inteighty in the guest, the kernel's handler can check if the instruction came from the user-space or from an external source by examining the previously executed instruction referenced by the RIP on the context stack or by checking the APIC page.
For other interrupts such as \intzero, determining if it is a genuine or a \spurious interrupt is unclear because it would require analyzing and interpreting the executed user space code.

\paragraph{Disabling Interrupt Handlers.}
Another approach is to disable vulnerable interrupts by not registering handlers for them in the guest OS. 
This works for \inteighty if the kernel is recompiled without the configuration flag \verb|CONFIG_IA32_EMULATION|, which disables IA32 emulation.
However, again, this does not generalize beyond \inteighty; and even then may break compatibility with legacy code that relies on \inteighty behavior. 
We survey 5 flavors of GCP- and Azure-recommended CVM images (Redhat, Fedora, CentOS, Ubuntu), standalone Debian-rolling, and ArchLinux.
All of them have kernels with 32-bit support compiled in at the time of writing.
It is required to ensure maximal compatibility and guarantee legacy support.
Linux 6.6. onwards it is possible to dynamically disable \verb|CONFIG_IA32_EMULATION| at boot time. 

\paragraph{TDX Implementation.}
For Intel TDX, we implemented both software-based defenses. 
First, we compiled the Linux kernel with the \verb|CONFIG_IA32_EMULATION| flag disabled in the configuration.
Second, detecting if an int 0x80 came externally required a patch of 14~\loc
where we checked the APIC page bit. Since this is the only way to inject external interrupts on TDX, this patch was sufficient. 
When running a user application in the guest OS that did a genuine \inteighty, servicing it on our patched kernel resulted in an overhead of 460\, cycles when compared to a vanilla kernel. 
We tested \codename on both these patched versions on Intel TDX and confirmed that the attack does not go through. 

We co-operated with Intel and Linux kernel developers to apply the second approach that detects external interrupts to protect TDX VMs against \codename. 
By default, TDX VMs execute with the IA32 emulation enabled and a patch to the guest kernel checks the APIC page bit to stop external injections of int 0x80~\cite{int80-path-2}. 

\paragraph{SEV-SNP Implementation.}
We attempted to implement the defense of detecting interrupts by examining the APIC page.
On AMD, the hypervisor can inject external interrupts asynchronously via the APIC page (same as TDX).
Similar to TDX, we implemented the virtual APIC check for AMD SEV-SNP with 14\,LoC. 
We observed an overhead of 10182\,cycles compared to the original unpatched execution of a binary that genuinely performs \inteighty. 
However, this is insufficient on AMD because the hypervisor can also inject interrupts via the VMCB registers which are handled when the VM resumes execution~(\Cref{sec:inject-interrupts}).
To stop this attack surface we used the defense strategy of detecting external interrupts by examining the last instruction that the user-space application executed. 
This requires examining the memory referenced by the \rip on the saved context stack to check if the program executed an \texttt{int} instruction with \texttt{0x80} as a parameter.
This requires disassembling the \rip in reverse for 2-bytes (since \inteighty results in a 2-byte opcode), where we inevitably run into classic problems stemming from variable length instructions. Determining if the user-code indeed performed \inteighty or some other stream of instructions and parameters that result in the same opcodes is undecidable. 
Thus our patches provide incomplete protection. 

To defend against \codename, the Linux kernel introduced a patch that disables IA32 emulation by default for SEV VMs~\cite{int80-path-1}. 
While this software patch stops \codename{}'s int 0x80 attacks, it is ineffective against attacks from interrupts (e.g., int 0x0) which are converted to signals. 
Detecting external injections of these interrupts using the \rip is not feasible. 
To decide if an interrupt is legitimate, the guest kernel would need to parse the whole instruction (opcode and all arguments), and in some cases emulate the instruction.
For example, to check if the application legitimately caused an overflow resulting in an int 0x10, the guest kernel would need to emulate the full arithmetic operation to reliably determine overflow conditions. 
Protecting against these interrupts would require hardware-based filtering techniques in~\cref{ssec:hw-based-filtering}.

\paragraph{Using Restricted \& Alternate Injection.}
We attempted to leverage the restricted and alternate mode to implement a software defense that adds the missing checks at least for \intzero and \inteighty. However, due to lack of software support for these modes in the hypervisor and the guest OS, we were unable to prototype these checks. 
One can implement stand-alone restricted injection directly in the host Linux kernel. However, there are no open-source implementations that we can test. 
Further, an initial patchset proposed by Microsoft received strong pushback by the Linux community~\cite{lkml-discussion}.
The main criticism for rejecting the patches was that a nested \#HV might corrupt the stack and hardware cannot protect against this race condition.
One can also implement restricted and alternate mode in combination, which necessitates nested virtualization to take advantage of VMPLs. 
Prior works that implement such nested virtualization for AMD SEV report 
high performance cost---throughput drops between 57\% and 85\% for MySQL, memcached, and Nginx~\cite{hecate}.
We anticipate further slowdown for interrupt filtering since, for each interrupt injection the host has to schedule the vCPU running in VMPL0 followed by the vCPU in a higher VMPL running the nested guest Linux OS.

\spacesave
\subsection{ Hardware-based Selective Filtering}
\label{ssec:hw-based-filtering}
Instead of relying on kernel patches that may break compatibility, hardware-level filtering offers a cleaner defense. 
One extreme solution is to filter all external interrupts for the CVM, but this breaks critical functionality such as timers. Instead, we propose selective filtering of interrupts that typically have explicit effect handlers.

\paragraph{TDX.}
Intel already blocks interrupts 0-31 from APIC by default. 
If the hypervisor needs to inject necessary interrupts between 0-31 (e.g., NMI), 
it needs to use the TDX interface. The trust domains module (TD module), which is in the TCB, provides this interface and determines whether to forward it to the CVM. As we reported in \cref{sec:handler-results}, none of the interrupts between 0-31 with explicit effect handlers are forwarded by the TD module. 
We recommend that TDX should treat \inteighty the same as 0-31 and filter it. This will break legacy code that may externally inject \inteighty~\cite{xen-int80, inria-int80}.

\paragraph{SEV-SNP.}
We recommend that SEV should employ similar filtering of all externally injected interrupts that may have explicit effect handlers. 
Doing such filtering in microcode can provide comprehensive protection against \codename.
While the same effect can perhaps be achieved with the restricted and alternate modes, we have two reservations. 
This requires correctly patching several codebases for hypervisors, guest OSes, and VMPL0 implementations. 
Since we were not able to test the complete and functional implementations of these modes, it is unclear if they are completely robust against hypervisors.
Specifically, one needs to ensure that the hypervisor has no way to: (i) inject these interrupts via the APIC or the synchronous interface; (ii) disable the restricted and alternate modes at any point during the CVM's execution; (iii) re-enter the handlers to exploit race-conditions or break atomicity and nested interrupt assumptions~\cite{google-tdx-report}. 
The upcoming secure AVIC proposal from AMD is a good candidate to achieve hardware-level filtering, 
where the CVM can specify a hardware interrupt filter without software intervention~\cite{secure-avic}.

\spacesave
\section{Related Work}

Previous works attack SEV's memory protection to inject arbitrary code to the CVM.~\cite{236278, SEVurity}
CrossLine attacks use hypervisor-controlled address space identifiers (ASIDs) to compromise SEV VMs just before they crash~\cite{Crossline}.  
Further, there have been numerous exploits that compromise SEV VMs using side-channels~\cite{Cipherleaks, 9833768, SEVerESt, SEVered}.
Buhren~\etal~\cite{10.1145/3319535.3354216} %
compromise SEV's remote attestation mechanisms to extract platform keys and perform arbitrary code injection in SEV VMs. 
Zhang~\etal architecturally revert modified cache lines to break SEV~\cite{cachewarp}.
Buhren~\etal mount fault injection attacks against SEV-SNP VMs by extracting endorsement keys using voltage glitching~\cite{Glitch}.
SEV-ES has been shown to offer much weaker security than SEV-SNP~\cite{sev-snp}. 
However, \codename breaks SEV-SNP guarantees without relying on any micro-architectural, architectural, power, or glitching side-channels.
Google performed a security review of Intel TDX and SEV SNP and reported several issues~\cite{google-tdx-report,google-sev-vuln}. 
Notably, on TDX they found a vulnerability that allowed untrusted firmware to induce software exceptions during the early boot stages.
Using this, they gain control over the instruction pointer during trusted firmware execution, thus achieving arbitrary code execution.
To the best of our knowledge, \codename is the first attack on TDX 
from untrusted hypervisor. Further, we do not control the instruction pointer, instead we re-use the handlers in the trusted software (guest OS and user applications). 
AMD emphasizes that the hypervisor must respect RFLAGS.IF to preserve guest kernel functionality~\cite{sev-snp}, but 
\codename does not violate this flag. Future works can explore the combination of \codename with this mechanism to exploit the      kernel~\cite{exprace}.

\paragraph{Tooling.}
SEV-step and SGX-step use timer interrupts to build single-stepping primitives for SEV-SNP VMs and SGX enclaves respectively~\cite{sev-step,sgx-step, cachewarp}.
\codename does not require the full-fledged suite of primitives offered by these tools and they do not apply out-of-box for our attack. However, when we build our tooling, we re-use valuable insights and implementation details from these tools.

\paragraph{Lift and Shift.}
Porting legacy applications to  TEE platforms with zero developer efforts is referred to as lift-and-shift. 
Porting applications to Intel SGX entails maintaining  compatibility~\cite{haven, graphene-sgx, occlum, scone} and performance~\cite{scone, occlum}.
CVMs, due to their VM abstraction, reduce the overheads of porting legacy applications. 
However, using AMD SEV-SNP and Intel TDX still requires enlightening the guest OS to ensure that legacy code written with the assumption of a trusted hypervisor is protected in the TEE threat model. 
Further, the untrusted hypervisor also needs to support the creation of CVMs for different TEE backends. 
To this end, Intel, AMD, and several hypervisor solutions such as KVM and Hyper-V are working towards patching the hypervisors and guest OSes. 
Other approaches introduce a trusted manager inside the CVM that acts as a bridge between the hypervisor and the guest OS, removing the need to patch existing guest OSes. Recent works have shown that one can leverage AMD SEV-SNP's VMPL modes to achieve this goal~\cite{hecate}.
All of these works emphasize and aim to protect against the threats of untrusted privileged software. However, their reasoning about malicious interrupts, especially for CVMs, is either missing or incomplete.

\paragraph{Interface Security.}
Previous works that attack Intel SGX enclaves show the importance of correctly securing untrusted interfaces (e.g., system calls)~\cite{checkoway2013iago, coin}. 
Several works exploit interfaces of various TEEs to leak secret keys and enable remote code reuse~\cite{taleoftwoworlds, controlled-data-race, boomerang, hpe}. 
In a similar vein, \codename abuses the interrupt interface controlled by the untrusted hypervisor but for CVMs which offer a different abstraction.

\paragraph{Physical vs. Virtual Interrupts.}
Physical interrupts, including timers and page faults, are transparent---the victim application/CVM does not recognize it was interrupted and resumed. This allows the attacker to observe side-effects of said interruption~\cite{sev-step,sgx-step}. Defenses such as  AEX-Notify make the victim aware of physical interrupts, such that it can take preventive actions~\cite{aex-notify}.
\codename observes that virtual interrupts are not transparent to the CVM, they do not cause a VM exit but instead the CVM actively reacts to them as if they were benign interrupts.
One effect of such unexpected virtual interrupts is that the victim VM crashes (e.g., invalid opcode in kernel mode) or resumes execution (e.g., timers). This can perhaps be used to amplify side-channels, as is the case with physical interrupts. 
More importantly, \codename shows that certain virtual interrupts, when injected at the right time and location, have explicit effects that alter the register state of the victim CVM. 
\codename is the first work that abuses the virtual interrupt injection interface to alter the guest state to break the execution integrity of CVMs.
WeSee~\cite{wesee-oakland} is our follow-up work on \codename. It expands our analysis but focuses on one particular interrupt vector 29, VMM communication exception (\#VC), which was introduced in AMD SEV-SNP. Refer to  Appendix~\ref{appx:intr-classification} for further details.

\paragraph{Interrupt Protection.}
Wojtczuk and Rutkowska showed that in a mutually untrusted co-tenant VM setting, attackers can use rogue devices to perform interrupt injection attacks~\cite{msi-inj-attack}. Next, we discuss prior works that focus on TEE settings. 
Isolated computation on low-end micro-controllers can be made resistant to interrupt/exception attacks (e.g., timer interrupts for side-channels) with 
programming mechanisms~\cite{6868649, 10.1145/3470534, 9155211}.
TrustZone's secure interrupts can isolate interrupts of the secure-world from the untrusted normal world~\cite{TZOS}. 
AEX-Notify makes SGX enclaves aware of timer interrupt~\cite{aex-notify} using an ISA extension. Specifically, enclaves can register interrupt handlers to thwart single-stepping attacks stemming from timer interrupts.

\paragraph{Arm CCA.}
Unlike x86, Arm uses different interrupt architecture and nomenclature.
The Arm defines 4 classes of exceptions (synchronous exception, IRQ, FIQ, and SError).
We study the Arm CCA support for creating CVMs and report that it only allows injection of IRQs and FIQs into Arm CCA CVMs. The rest are filtered by the trusted Realm Management Monitor (RMM).
We tested all the IRQs and FIQs with RMM v0.3.0 and did not observe explicit effect handlers. 
Arm does not have a concept of a syscall interrupt like x86.

\spacesave
\section{Conclusion}
\codename presents a new attack on Intel TDX and AMD SEV-SNP that offer VM abstractions. 
It uses the untrusted hypervisor's interrupt management and delivery interface to inject malicious interrupts into CVMs.
\codename's gadgets use the explicit and global effects of the interrupt handlers to change the data and control flow of victim programs.
By injecting particular malicious interrupts at the right time in the right core, \codename breaks the integrity and subsequently confidentiality of CVM.
Our case-studies show the severity of \codename and highlight the need for robust defenses.

\section*{Acknowledgement}
We thank our shepherd, the anonymous reviewers, and Mélisande Zonta-Roudes for their constructive feedback for improving the paper. Thanks to Intel, AMD, and Linux for the mitigation discussions and for developing the patches.
We thank Benny Fuhry and Mona Vij from Intel for granting us early-access to TDX pre-production machines.

\bibliographystyle{plain}
\bibliography{references} 

\appendix

\section{Interrupt Injection Flow}
\label{appendix:injection}
\begin{lstlisting}[language=c,label={lst:example},caption={\openssh mm\_anwser\_auth\_password function}]
<mm_anwser_auth_password>
...
call auth_password
test eax,eax
...
\end{lstlisting}

\begin{enumerate}
    \item Guest executes \texttt{ret} in \authpasswd
    \item CPU fetches page of \mmanspwd to continue execution at line 3 in \mmanspwd.
    \item CPU throws a stage-2 page fault, since the page which contains \mmanspwd is marked as non-executable.
    \item CPU exits VM mode and transfers control back to the hypervisor.
    \item Hypervisor clears NX (non-executable) bit of \mmanspwd stage-2 entry
    \item Hypervisor modifies \texttt{VMCB.EventInj} field to inject interrupt into VM
    \item Hypervisor executes \texttt{VMRUN}
    \item \texttt{VMRUN} evaluates \texttt{VMCB.EventInj} field and signals an Interrupt before guest execution is resumed.
    \item Guest Kernel handles interrupt (int0x80)
    \item Guest continues execution on line 3 with corrupted state
\end{enumerate}
Technically this corresponds to a window of single instruction from the victim CVM point of view. However, our page fault mechanism exits the VM. Then on the hypervisor, the attacker can take as much time as it wants to set the interrupt vector and resume the VM. On resumption, subsequent steps happen out of the box.
\section{Interrupt Injection Interfaces}
AMD has four different interrupt injection interfaces implemented in their virtualization extensions as summarized in~\cref{tab:sev-interfaces}. 
As described in the main text we exclusively used \texttt{VMCB.EventInj} in our proof-of-concept implementation.
It allows to serve interrupts when a CVM is resumed with the \texttt{VMRUN} instruction.
Next, we have the virtual APIC (AVIC Advanced Virtual Interrupt Controller in AMDs nomenclature).
This is used for asynchronous interrupt injection, i.e., the vCPU does not need to exit and enter again to receive an interrupt.
We did not use the \texttt{V\_IRQ} but the documentation indicates it does the same as the \texttt{VMCB.EventInj}. 
Most interestingly is the last interface, the Physical IRQ interception bit. 
We can unset this bit on \texttt{VMRUN} and cause all physical interrupts to be received by the CVM rather than causing a \texttt{VMEXIT}. 
However, this bit is ignored when Restricted / Alternate Injection is enabled. 
Thus this interface cannot be used to circumvent those modes.
TDX has a similar interface but is not susceptible. 
The interrupt interception bit is controlled by the TD module that is part of the TCB.
\begin{table}[]
\caption{SEV-SNP interfaces for interrupt injection to VM~\cite{amd-manual}}
\vspace{9pt}
\label{tab:sev-interfaces}
\resizebox{\columnwidth}{!}{%
\begin{tabular}{lll}
\hline
\multicolumn{1}{c}{AMD VM interface} & \multicolumn{1}{c}{Injection Point} & \multicolumn{1}{c}{Effect} \\ \hline
\begin{tabular}[c]{@{}l@{}}VMCB \\ Event Injection\end{tabular} &
  \begin{tabular}[c]{@{}l@{}}synchronous\end{tabular} &
  \begin{tabular}[c]{@{}l@{}}raise interrupt before the first guest\\ instruction is executed\end{tabular} \\ \hline
virtual APIC &
  \begin{tabular}[c]{@{}l@{}}asynchronous\end{tabular} &
  \begin{tabular}[c]{@{}l@{}}raise interrupt whenever the AVIC \\ registers a change in the IRR register\end{tabular} \\ \hline
\begin{tabular}[c]{@{}l@{}}VMCB \\ V\_IRQ\end{tabular} &
  \begin{tabular}[c]{@{}l@{}}synchronous\end{tabular} &
  \begin{tabular}[c]{@{}l@{}}VMRUN loads the intr. information \\ into the respective on-chip registers\end{tabular} \\ \hline
\begin{tabular}[c]{@{}l@{}}Physical IRQ \\Intercept\end{tabular} &
  \begin{tabular}[c]{@{}l@{}}asynchronous\end{tabular} &
  \begin{tabular}[c]{@{}l@{}}Physical interrupts are interpreted as\\virtual interrupts and do no exit the VM\end{tabular} \\ \hline
\end{tabular}%
}
\end{table}

\section{Busy Wait Loop}
The code busy loops until \texttt{authenticated} becomes equal to one. This is used in \openssh and sudo for the TDX proof-of-concept exploits.
\begin{lstlisting}[language=c,label={lst:busyloop}]
// assume authenticated == 0
__asm__ __volatile__("1:;\n"
					"nop\n"
					"test %
					"je 1b \n"		
					:"+a"(authenticated):);
return authenticated;
\end{lstlisting}

\section{Tracing Overhead}
\label{appx:trace:overhead}

Table~\ref{tab:trace:overhead} shows the time taken to generate boot trace, application set (\sapp), and page trace (\tapp), when compared to the execution of the CVM without page fault tracing.
During the offline phase, we get multiple traces as summarized in Table~\ref{tab:traces}. In this case, the overheads of tracing slowdown the function generation described in Section~\ref{sec:inject-interrupts}. While this can be further optimized, we did not put efforts in such optimizations since this is a preparatory step before the victim runs its VM.

\codename also enables page fault tracing in the online phase, i.e. 
when the victim starts interacting with the VM. 
\codename causes some slowdown but it does not impact the victim's usability or result in detection. This is because for \openssh and sudo, we can potentially perform the tracing and the injection after attestation, but during the CVM provisioning which can take several minutes even in a benign setting. Thus, \codename attack happens before the user gets access to the VM, so it will not notice the lag. In cases where this is not possible (e.g. MLP), we can further cap the number of page faults for any given page (max capture in~\Cref{tab:traces}) to reduce the lag.

\begin{table*}
\centering
\centering
\small
\caption{\centering Overheads to generate boot trace, application set (\sapp), and page trace (\tapp) w.r.t. execution without page faults.}
\spacingtable
\label{tab:trace:overhead}
\begin{tabular}{lrrrrrrrrr} 

\toprule

 App       & \multicolumn{3}{c}{boot}                                                                                             & \multicolumn{3}{c}{\sapp}                                                                                              & \multicolumn{3}{c}{\tapp}                                                                                              \\ 
\hline
        & \multicolumn{1}{c}{\begin{tabular}[c]{@{}c@{}}trace\\off\end{tabular}} & \multicolumn{1}{c}{\begin{tabular}[c]{@{}c@{}}trace\\on\end{tabular}} & \multicolumn{1}{c}{+\%} & \multicolumn{1}{c}{\begin{tabular}[c]{@{}c@{}}trace\\off\end{tabular}} & \multicolumn{1}{c}{\begin{tabular}[c]{@{}c@{}}trace\\on\end{tabular}} & \multicolumn{1}{c}{+\%} & \multicolumn{1}{c}{\begin{tabular}[c]{@{}c@{}}trace\\off\end{tabular}} & \multicolumn{1}{c}{\begin{tabular}[c]{@{}c@{}}trace\\on\end{tabular}} & \multicolumn{1}{c}{+\%}  \\ 
\toprule

\noalign{\vskip 2mm}   

\openssh & 10.01s                                                             & 11.41s                 & 14\%                    & 14.2ms                                                             & 32.9ms                 & 131\%                    & 14.2ms                                                             & 773.9ms                & 5332\%                   \\
Sudo    & 10.01s                                                             & 13.68s                 & 37\%                    & 1.00s                                                              & 1.03s                  & 3\%                      & 1.00s                                                              & 7.02s                  & 602\%                    \\

MLP     & 8.71s                                        & 12.01s                 & 38\%           &   1.08s                                                                 &       1.11s                 &    3\%                      &                                                       1.08             &    1.95s
&         81\%   

\\

        \bottomrule
\end{tabular} %
\end{table*}

\section{Interrupt Classification}
\label{appx:intr-classification}

\paragraph{Custom Interrupt Injections.} To find interrupts of interest, we analyze the impact of malicious injections on a guest VM kernel and userspace application. 
For this, we survey all 256 interrupt vectors that are supported by current AMD SEV-SNP-enabled CPUs. 
To get started, we examine the legacy interrupt injection functions in KVM that are used by the hypervisor to inject benign interrupts. 
On the host kernel, we hook these KVM functions to build a kernel module interface. 
Using a custom kernel module, we can now selectively inject interrupts into VMs as explained in~\cref{sec:inject-interrupts}.

\begin{lstlisting}[language=c, caption={User-space application with general signal handler}, label={lst:signalcatcher}]
jmp_buf jumper;
void handler(int signum){
    printf("in sig handler\n");
    longjmp(jumper, 1);
}
int main()
{
    volatile int i, j;
    struct sigaction act;
    struct sigaction oldact;
    memset(&act, 0, sizeof(act));
    act.sa_handler = handler;
    act.sa_flags = SA_NODEFER | SA_NOMASK;
    for (i = 0; i <= 32; i++){
        if (sigaction(i, &act, NULL)){
            printf("Sig. install err\n");}}
    while (1){
        int x;
        asm("label2:");
        printf("Reached outside of loop\n");
        x = setjmp(jumper);
        if (x == 0){
            loop1:
            while (1){
                asm("cmpl $0, %
                    "jne label2");
            }
        } else {
           goto loop1;
        }
    }
    return 0;
}
\end{lstlisting}

\paragraph{Setup.} Similar to the end-to-end attacks in~\cref{sec:e2e-attacks}, we use a workstation with an EPYC 9124 CPU with 16 cores and 192\,GB RAM. We boot a Linux kernel v6.5.0-rc2 from AMD~\cite{AMDSEV-host-git} as the host operating system, which is modified to support our injection hooks. To start the VMs and perform our experiments, we use QEMU version 8.0.0. For the SEV-SNP VM we use the same Linux kernel as for the host, but without the patches. 
To test if the guest kernel generates a signal based on the injected interrupt, we implement a userspace application that is capable of catching these signals. The application (sources in Lst.~\ref{lst:signalcatcher}) registers dummy handlers for signals 0-31 with the kernel and busy-waits for their delivery. Furthermore, it checks continuously if registers, such as \verb|eax|, were modified, for example by the kernel.
To ensure that we are actually injecting the the interrupt while our application is executing, we run and pin it on each available vCPU (8 in our case) using \texttt{taskset}. This lowers the risk of the interrupt being delivered during kernel execution, by preventing it from idling.

\paragraph{Evaluation.} To classify the interrupts, we first manually analyzed the interrupt handler of the guest kernel for the anticipated behavior. This gives us the ability to compare the actual behavior of the OS to the injected interrupt with an expected normal execution flow. To the best of our abilities, we classify the predictions into the following categories using manual analysis:
\begin{enumerate}
    \item \textbf{Signal to usermode}: The kernel is expected to generate and send a signal to the user space application when the interrupt arrives while the core is executing in user space.
    \item \textbf{Specific kernel handler} The kernel has reserved handlers that are executed in the kernel when an interrupt of this type is delivered. However, it does not send any signal to user space.
    \item \textbf{Default kernel handler} The kernel has not registered any special handling for this type of interrupt. Instead, a basic handler is executed that logs the delivery and resumes the execution.
    \item \textbf{Kernel Panic} When an interrupt of this type is delivered, the kernel reacts with an unrecoverable panic and halts the execution on all vCPUs.
    \item \textbf{Syscall} We expect the kernel to handle a system call originated by the user space application.
    \item \textbf{Unclear} The expected behavior of this interrupt handler is inconclusive from our manual analysis.
\end{enumerate}

\begin{table*}[]
\caption{Summary of observations for injecting interrupts into a guest VM on AMD SEV-SNP (per vector).}
\vspace{9pt}
\label{tab:analysis}
\centering
\resizebox{0.9\textwidth}{!}{%
\begin{tabular}{@{}lllll@{}}
\toprule
Int. (Dec) & Int. (Hex) & Description                         & Expected behavior in guest VM & Observed behavior         \\ \midrule
0          & 00         & Divide by 0                         & Signal to usermode            & Signal to usermode (SIGFPE)        \\
1          & 01         & Debug                               & Signal to usermode            & Signal to usermode (SIGTRAP)       \\
2          & 02         & NMI                                 & Specific kernel handler       & VM\_EXIT fail             \\
3          & 03         & Breakpoint                          & Specific kernel handler       & VM\_EXIT fail            \\
4          & 04         & Overflow                            & Signal to usermode            & VM\_EXIT fail            \\
5          & 05         & Bound Range Exceeded                & Signal to usermode            & VM\_EXIT fail            \\
6          & 06         & Invalid Opcode                      & Signal to usermode            & Signal to usermode (SIGILL)       \\
7          & 07         & Device not available                & Specific kernel handler       & App crash                 \\
8          & 08         & Double Fault                        & Kernel Panic                  & Kernel panic              \\
9          & 09         & Co-Processor Segment overrun        & Signal to usermode            & VM\_EXIT fail            \\
10-13      & 0A-0D      & TSS/Segment/Protection Faults       & Unclear                       & App crash                 \\
14         & 0E         & Page Fault                          & Specific kernel handler       & Kernel panic              \\
15         & 0F         & Spurious Interrupt                  & Unclear                       & VM\_EXIT fail            \\
16         & 10         & x87 Floating Point Exception        & Unclear                       & No effect                 \\
17         & 11         & Alignment Check                     & Specific kernel handler       & App crash                 \\
18         & 12         & Machine Check                       & Specific kernel handler       & No effect                 \\
19         & 13         & SIMD Floating Point Exception       & Unclear                       & No effect                 \\
20         & 14         & Virtualization Exception            & Specific kernel handler       & VM\_EXIT fail            \\
21         & 15         & Control Protection Exception        & Specific kernel handler       & App crash                 \\
22-28      & 16-1C      & Undefined                           & Unclear                       & VM\_EXIT fail            \\
29         & 1D         & VMM Communication Exception         & Specific kernel handler       & VM\_EXIT fail            \\
30         & 1E         & Undefined                           & Unclear                       & Kernel panic              \\
31         & 1F         & Undefined                           & Unclear                       & VM\_EXIT fail            \\
32         & 20         & IRET Exception                      & Specific kernel handler       & No effect                 \\
33         & 21         & Undefined                           & Default kernel handler        & Default handler executed  \\
34         & 22         & Undefined                           & Default kernel handler        & No effect                 \\
35-47      & 23-2F      & Undefined                           & Default kernel handler        & Default handler executed  \\
48         & 30         & ISA IRQ                             & Specific kernel handler       & No effect                 \\
49         & 31         & ISA IRQ                             & Specific kernel handler       & Default handler executed  \\
50         & 32         & ISA IRQ                             & Specific kernel handler       & System unresponsive       \\
51-63      & 33-3F      & ISA IRQ                             & Specific kernel handler       & Default handler executed  \\
64-127     & 40-7F      & Undefined                           & Default kernel handler        & Default handler executed  \\
128        & 80         & Syscall                             & Syscall                       & App control flow changed  \\
129-235    & 81-EB      & Undefined                           & Default kernel handler        & Default handler executed  \\
236-243    & EC-F3      & Local Timer and Hypervisor Int.     & Specific kernel handler       & No effect                 \\
244        & F4         & Deferred Error                      & Specific kernel handler       & Specific handler executed \\
245-247    & F5-F7      & IRQ Work + x86 IPI Interrupts       & Specific kernel handler       & No effect                 \\
248        & F8         & Reboot Interrupt                    & Specific kernel handler       & System unresponsive       \\
249-250    & F9-FA      & Threshold + Thermal APIC Int.       & Specific kernel handler       & Specific handler executed \\
251-254    & FB-FE      & Function Call, Resched., Error Int. & Specific kernel handler       & No effect                 \\
255        & FF         & Spurious APIC Interrupt             & Specific kernel handler       & System unresponsive       \\ \bottomrule
\end{tabular}%
}
\end{table*}
\twocolumn

We classify the observable behavior into the following categories:
\begin{enumerate}
    \item \textbf{Signal to user space (TYPE)} We observed that a TYPE signal was received in the user space application
    \item \textbf{Specific handler executed} The guest VM executed a dedicated interrupt handler for this vector in the kernel.
    \item \textbf{Default handler executed} The guest VM executed the basic placeholder interrupt handler without any implications.
    \item \textbf{No effect} The invocation of the interrupt did not have an observable effect on the guest VM.
    \item \textbf{System unresponsive} The status of the guest VM is inconclusive; no information exchange with the guest Kernel is possible after the interrupt injection.
    \item \textbf{App crash} The execution of our user space application was terminated, but the OS was able to continue operating normally.
    \item \textbf{App control flow changed} We were able to see an impact on the control flow of the executed user space application.
    \item \textbf{VM\_EXIT fail} After injecting the interrupt, the hypervisor was not able to continue the execution of the VM as the CPU refused to enter the guest successfully.
\end{enumerate}
We present a summary of our findings in~\cref{tab:analysis}.
We used the \texttt{svm\_deliver\_interrupt} function within the Linux kernel to inject all the interrupts. 
This API does not allow us to set an error code nor does it always set the right hardware flags for all interrupts (i.e., NMIs are supposed to be injected using a different API).
Thus, for all interrupts where we report {\em VM\_EXIT fail}, one can still inject these interrupts in the VM by directly modifying the respective fields in the \texttt{VMCB}.
Furthermore, some interrupts might panic the kernel only under certain circumstances and have different effects if the interrupt is raised at the right time (e.g., int3 on a debug instruction).
As a concrete example, we consider interrupt 29 which was flagged as ``VM\_EXIT fail'' in our analysis. On further manual investigation, we identified that by directly manipulating \texttt{VMCB.EventIn}, we can inject the interrupt into the VM and cause a termination. By further modifying the \texttt{error\_code} we can circumvent the crash and cause different global state change effects when the CVM resumes execution.
We investigate the impact of this behavior in detail in WeSee~\cite{wesee-oakland}.
For the scope of \codename, we do not do an extensive analysis of: (a) all potential 
hardware features and interfaces to inject interrupts and (b) all injection points during the victim execution. We leave this analysis as well as effects of combining (a) and (b) to detect other instances of Ahoi attacks to future work.

\end{document}